\documentclass[aps,twocolumn,groupedaddress]{revtex4}

\begin{document}
\bibliographystyle{apsrev}

\title{An Introduction to Quantum Order, String-net Condensation,\\
and Emergence of Light and 
Fermions\footnote{A more comprehensive description of
topological/quantum orders can be found in 
\emph{Quantum field theory of many-body
systems -- from the origin of sound to an origin of light and fermions},
Xiao-Gang Wen, Oxford Univ. Press, 2004.}}

\author{Xiao-Gang Wen}
\homepage{http://dao.mit.edu/~wen}
\affiliation{Department of Physics, Massachusetts Institute of Technology,
Cambridge, Massachusetts 02139}

\date{Jan. 2004}

\begin{abstract}
We review some recent work on new states of matter. Those states 
cannot be described symmetry breaking and hence contain a new
kind of order -- quantum order.
Some quantum orders are shown to be closely related to string-net
condensations. Those quantum orders lead to an emergence of gauge bosons and
fermions from pure bosonic models.
\end{abstract}
\keywords{Quantum orders, Gauge theory, String-net condensation}

\maketitle

\section{Introduction}

\subsection{Origin of light/fermion and new orders}

The existences of light and fermions are two big mysteries in nature.  The
mysteries are so deep that the questions like, ``What are light and
fermions?'', ``Where do light and fermions come from?'', ``Why do light and
fermions exist?'', are regarded by many people as philosophical or even
religious questions.  

To appreciate the physical significance those questions let us ask three
simpler questions: ``What are phonons?'', ``Where do phonons come from?'',
``Why do phonons exist?''.  We know these three questions to be scientific
questions and we know their answers. Phonons are vibrations of a crystal.
Phonons come from a spontaneous translation symmetry breaking.  Phonon exists
because the translation-symmetry-breaking phase actually exists in nature.  It
is quite interesting to see that our understanding of a gapless excitation --
phonon -- is rooted in our understanding of the phases of matter as symmetry
breaking states \cite{N6080,G6154}.  

However, our picture for massless photons and nearly massless
fermions\footnote{All the observed particles can be treated as massless when
compared with Planck mass.} (such as electrons and quarks) is quite different
from our picture of gapless phonons. We regard photons and fermions as
elementary particles -- the building block of our universe.

But why should we regard photons and fermions as elementary particles?
Why don't we regard photons and fermions as emergent quasiparticles like
phonons? We can view this question from several different angles.

\noindent \textbf{First point of view}:
Before late 1970's, we felt that we understood, at least in principle, all the
physics about phases and phase transitions based on Landau's symmetry breaking
theory \cite{L3726,LanL58}.  In such a theory, if we start with a bosonic
model, the only way to get gapless excitations is via spontaneous breaking of
a continuous symmetry \cite{N6080,G6154}, which will lead to gapless
\emph{scalar bosonic} excitations. It seems that there is no way to obtain
gapless gauge bosons and fermions from symmetry breaking. This may be the
reason why people think our vacuum (with massless gauge bosons and
nearly-gapless fermions) is very different from bosonic many-body systems
(which were believed to contain only gapless scalar bosonic collective
excitations, such as phonons).  It seems there does not exist any order that
give rise to massless photons and nearly-massless fermions.  This may be the
reason why we regard photons and fermions as elementary particles and
introduce them by hand into our theory of nature.

\noindent \textbf{Second point of view}:
On the other hand, the resemblance between the photons and the phonons makes
it odd to regard photons as elementary.  To appreciate this point, let us
imagine another universe which contains three types of massless excitations.
These massless excitations behaves in every way like the phonons in a
crystal.  We will not hesitate to declare that the vacuum in that universe is
actually a crystal even when no one can see the particles that form the
crystal. Our conviction of the existence of the crystal does not come from
seeing the lattice structure, but from seeing the low energy collective modes
of the crystal.

Now back to our universe. Are the massless photons and nearly massless
fermions also collective modes of certain order in our vacuum.  Not knowing
what order can give rise to photons and fermions may not imply the photons and
fermions to be elementary. More likely, it means that our understanding of
order is incomplete.  The very existence of light and fermions may indicate
that our vacuum contain a new kind of order.  The new order will produce
light and fermions, and protect its masslessness.

\noindent \textbf{Third point of view}:
If we had a material which is described by bosons (such as a spin system) and
if we found that the low energy excitations in the material are gauge bosons
and fermions, we would not hesitate to declare that the material contains a
new kind of order beyond the symmetry breaking description.  But so far, we
have not find any material that contain emergent gauge bosons and emergent
fermions. So we do not know if new order beyond the symmetry breaking exists
or not. Other other hand, we may regard our vacuum as a special material. From
this point of view, the light and the electrons in the vacuum provided an
experimental evidence of the existence of new order.

\subsection{Topological order and quantum order}

Historically, our understanding of new order beyond symmetry breaking starts
at an unexpected place --- fractional quantum Hall (FQH) systems.  The FQH
states discovered in 1982 \cite{TSG8259,L8395} opened a new chapter of
condensed matter physics.  What is really new in FQH states is that FQH
systems contain many \emph{different} phases at zero temperature which have
the \emph{same} symmetry. Thus those phases cannot be distinguished by
symmetries and cannot be described by Landau's symmetry breaking theory.  

Since FQH states cannot be described by Landau's symmetry breaking theory, it
was proposed that FQH states contain a new kind of order --- topological order
\cite{Wtoprev}. Topological order is new because it cannot be described by
symmetry breaking, long range correlation, or local order parameters.  None of
the usual tools that we used to characterize a phase applies to topological
order. Despite this, topological order is not an empty concept since it can be
characterized by a new set of tools, such as the number of degenerate ground
states \cite{HR8529,WNtop}, the non-Abelian Berry's phase under modular
transformations \cite{Wrig}, quasiparticle statistics \cite{ASW8422}, and edge
states \cite{H8285,Wedgerev}.  Just like Ginzburg-Landau theory is the
effective theory of symmetry breaking order, the topological field theory
\cite{W8951} is the effective theory of topological order \cite{Wtoprev}.

It was shown that the ground state degeneracy of a topologically ordered state
is robust against \emph{any} perturbations \cite{WNtop}. Thus the ground state
degeneracy is a universal property that can be used to characterize a phase.
The existence of topologically degenerate ground states proves the existence
of topological order. The topologically degenerate ground states were found to
useful in fault tolerant quantum computing \cite{K032}.

\begin{figure}[tb]
\centerline{
\includegraphics[width=3.5in]{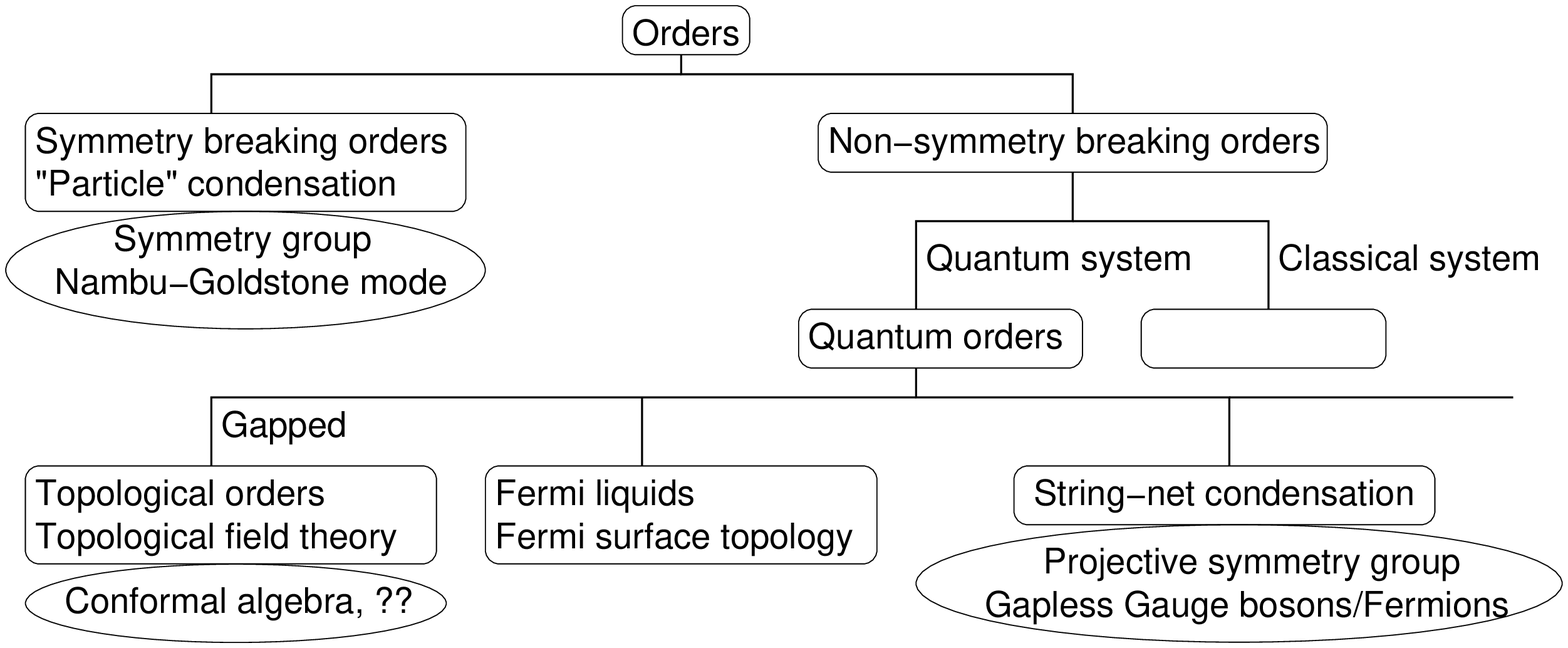}
}
\caption{
A classification of different orders in matter (and in vacuum).
}
\label{clorder}
\end{figure}

The concept of topological order only applies to state with finite energy gap.
It was recently generalized to quantum order \cite{Wqoslpub} to describe new
kind of orders in gapless quantum states.  There are two general but vague
ways to understand quantum orders.

In the first understanding, we assume that the order in a quantum state is
encoded in the many-body ground state wave function.  We believe that the
symmetry of the ground state wave function cannot characterize all the
possible orders in the many-body state.  The extra structure in the ground
state can be viewed as a pattern of quantum entanglement in the many-body 
state.  From this point of view, we may say that quantum orders are patterns
quantum entanglement in quantum many-body states.

The second way to understand quantum
order is to see how it fits into a general classification scheme of orders
(see Fig.  \ref{clorder}).  First, different orders can be divided into two
classes: symmetry breaking orders and non-symmetry breaking orders. The
symmetry breaking orders can be described by a local order parameter and can
be said to contain a condensation of point-like objects. The amplitude of
condensation corresponds to the order parameter. All the symmetry breaking
orders can be understood in terms of Landau's symmetry breaking theory. The
non-symmetry breaking orders cannot be described by symmetry breaking, nor by
the related local order parameters and long range correlations. Thus they are
a new kind of orders. If a quantum system (a state at zero temperature)
contains a non-symmetry breaking order, then the system is said to contain a
non-trivial quantum order. We see that a quantum order is simply a
non-symmetry breaking order in a quantum system.

Quantum orders can be further divided into many subclasses. If a quantum state
is gapped, then the corresponding quantum order will be called topological
order. 
The second class of quantum orders appear in Fermi liquids (or free fermion
systems). The different quantum orders in Fermi liquids are classified by the
Fermi surface topology \cite{L6030}. We will discuss this class of quantum
order briefly in section \ref{freeferm}. The third class of quantum orders
arises from a condensation of nets of strings (string-nets)
\cite{W7159,BMK7793,KS7595,Walight,LWsta,FNS0311,LWstrnet}.  We will discuss
it in sections \ref{prjcon} and \ref{str}.  This class of quantum orders
shares some similarities with the symmetry breaking orders of ``particle''
condensation.

We know that different symmetry breaking orders can be classified by symmetry
groups. Using group theory, we can classify all the 230 crystal orders in
three dimensions. The phase transitions between different symmetry breaking
orders are described by critical point with algebraic correlations.  The
symmetry also produces and protects gapless collective excitations -- the
Nambu-Goldstone bosons -- above the symmetry breaking ground state.
Similarly, different string-net condensations (and the corresponding quantum
orders) can be classified by a mathematical object called projective symmetry
group \cite{Wqoslpub} (see subsection \ref{psg}).  Using the projective
symmetry group, we can classify over 100 different 2D spin liquids that all
have the same symmetry.  The phase transitions between different quantum
orders are also described by critical points. Those phase transitions do not
change any symmetry and cannot be described by order parameters associated
with broken symmetries \cite{WWtran,CFW9349,SMF9945,RG0067,Wctpt}.  Just like
the symmetry group, the projective symmetry group can also produce and protect
gapless excitations.  However, unlike the symmetry group, the projective
symmetry group produces and protects gapless gauge bosons and fermions
\cite{Wqoslpub,Wlight,WZqoind}. Because of this, we can say that light and
massless fermions can have a unified origin.  They can emerge from string-net
condensations.

\subsection{String-net picture of light and fermions}

We used to believe that to have light and fermions in our theory, we have to
introduce by hand a fundamental $U(1)$ gauge field and anti-commuting fermion
fields, since at that time we did not know any collective modes that behave
like gauge bosons and fermions.  However, due to the advances of the last 20
years, we now know how to construct \emph{local bosonic systems} that have
emergent \emph{unconfined} gauge bosons and/or fermions
\cite{FNN8035,KL8795,WWZcsp,RS9173,Wsrvb,K032,MS0181,Wlight,MS0204,Walight,LWsta,MS0312,HFB0404}.
In particular, one can construct ugly bosonic spin models on a cubic lattice
whose low energy effective theory is the beautiful QED and QCD with emergent
photons, electrons, quarks, and gluons \cite{Wqoem}.

\begin{figure}[tb]
\centerline{
\includegraphics[width=2.5in]{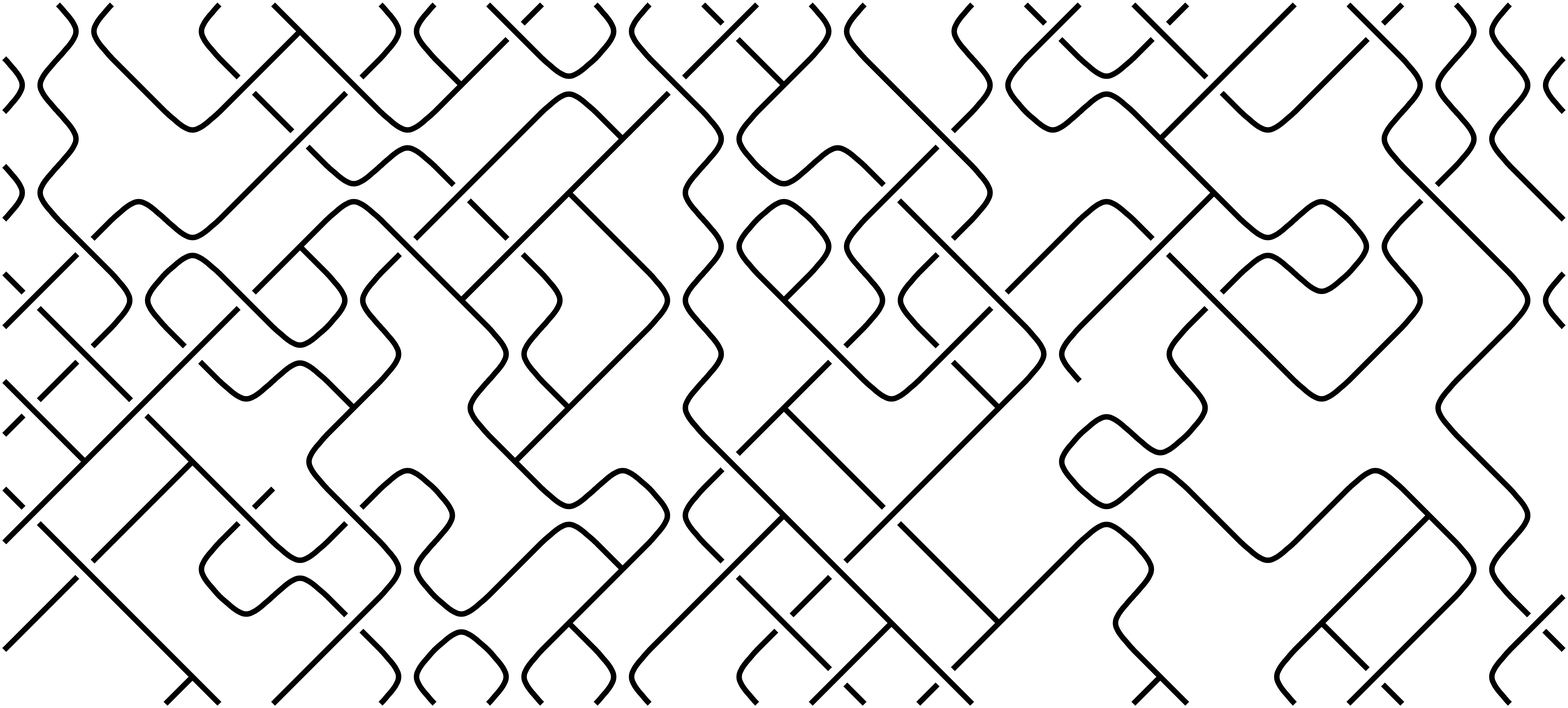}
}
\caption{
Our vacuum may be a state filled with string-nets. The fluctuations
of the string give rise to gauge bosons. The ends of the strings 
correspond to electrons, quarks, \etc.
}
\label{stringnetS}
\end{figure}

This raises an issue: do light and fermions in nature come from a fundamental
$U(1)$ gauge field and anti-commuting fields as in the $U(1)\times SU(2)
\times SU(3)$ standard model or do they come from a particular quantum order
in our vacuum?  Is Coulomb's law a fundamental law of nature or just an
emergent phenomenon?  Clearly it is more natural to assume light and fermions,
as well as the Coulomb's law, come from a quantum order in our vacuum.  From
the connections between string-net condensation, quantum order, and massless
gauge/fermion excitations, it is very tempting to propose the following
possible answers to the three fundamental questions about light and
fermions:\\
\textbf{What are light and fermions?} \\
Light is the fluctuation of condensed strings (of arbitrary sizes) 
\cite{KS7595,BMK7793,FNN8035}.  Fermions are ends of condensed 
strings \cite{LWsta}.\\
\textbf{Where do light and fermions come from?}  \\
Light and fermions come from the collective motions of nets of
strings (or string-net) that fill our vacuum (see Fig.
\ref{stringnetS}).\\
\textbf{Why do light and fermions exist?}   \\
Light and fermions exist because our vacuum happen to have a property called
string-net condensation.

Had our vacuum chose to have ``particle'' condensation, there would be only
Nambu-Goldstone bosons at low energies. Such a universe would be very boring.
String condensation and the resulting light and fermions provide a much
more interesting universe, at least interesting enough to support intelligent
life to study the origin of light and fermions.

Our understanding of quantum/topological orders are base on many researchs in
three main areas: (1) the study of topological phases in condensed matter
systems such as FQH systems \cite{WNtop,BWkmat2,R9002,FK9169}, quantum dimer
models \cite{RK8876,RC8933,MS0181,MS0312,AFF0493}, quantum spin models
\cite{KL8795,WWZcsp,Wrig,RS9173,Wsrvb,SF0050,Wqoslpub,SP0258,BFG0212}, or even
superconducting states \cite{Wtopcs,HOS0427}, (2) the study of lattice gauge
theory \cite{W7159,BMK7793,KS7595,K7959}, and (3) the study of quantum
computing by anyons \cite{K032,IFI0203,FLZ0205}.  In this paper, we will use
some simple models to introduce the main points of topological/quantum order.

\section{State of matter and concept of order}
\label{tr:0}

\begin{figure}[tb]
\centerline{
\includegraphics[width=1.2in]{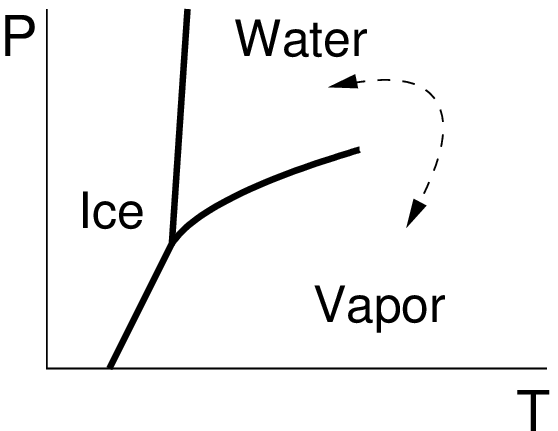}
}
\caption{
The phase diagram of water.
}
\label{waterphase}
\end{figure}

To start our journey to search new state of matter with emergent gauge bosons
and fermions, we like to first discuss the concept of order and review the
symmetry breaking description of order.  With low temperature technology
developed around 1900, physicists discovered many new states of matter (such
as superconductors and superfluids). Those different states have different
internal structures, which are called different kinds of orders. The precise
definition of order involves phase transition.  Two states of many-body
systems have the same order if we can smoothly change one state into the other
(by smoothly changing the Hamiltonian) without encounter a phase transition
(\ie without encounter a singularity in the free energy). If there is no way
to  change one state into the other without a phase transition, than the two
states will have different orders.  We note that our definition of order is a
definition of equivalent class.  Two states that can be connected without a
phase transition are defined to be equivalent. The equivalent class defined
this way is called the universality class. Two states with different orders
can be also be said as two states belong to different universality classes.
According to our definition, water and ice have different orders while water
and vapor have the same order. (See Fig. \ref{waterphase})
\index{phase}
\index{phase transition}
\index{universality class}

\begin{figure}[tb]
\centerline{
\includegraphics[height=0.7in]{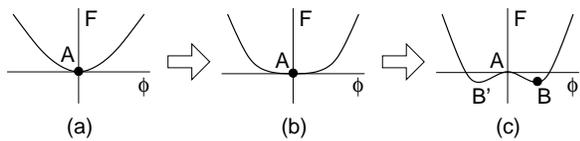}
}
\caption{
In the presence of $\phi\to-\phi$ symmetry, switching of the minima can be
continuous and causes a second-order phase transition. (a) The ground state is
symmetric under the transformation $\phi\to-\phi$. (c) The ground state breaks
the $\phi\to-\phi$ symmetry.  }
\label{scndphs}
\end{figure}

After discovering so many different kinds of orders, a general theory is
needed to gain a deeper understanding of states of matter.  In particular, we
like to understand what make two orders really different so that we cannot
change one order into the other without encounter a phase transition.  
It is a deep insight to connect the 
singularity in free energy to a symmetry breaking picture in Fig.
\ref{scndphs}.  Based on the relation between orders and symmetries, Landau
developed a general theory of orders and phase transitions
\cite{LanL58,GL5064}.  According to Landau's theory, the states in the
same phase always have the same symmetry and the states in different phases
always have different symmetries.  So symmetry is a universal property that
characterized different phases.  Landau's theory is very successful.
\index{symmetry breaking} Using Landau's theory and the related group theory for
symmetries, we can classify all the 230 different kinds of crystals that can
exist in three dimensions.  By determining how symmetry changes across a
continuous phase transition, we can obtain the critical properties of the
phase transition.  The symmetry breaking also provides the origin of many
gapless excitations, such as phonons, spin waves, \etc, which determine the
low energy properties of many systems \cite{N6080,G6154}. A lot of the
properties of those excitations, including their gaplessness, are directly
determined by the symmetry.  

\section{Quantum orders and quantum transitions in free fermion systems}
\label{freeferm}

However, not all orders are described by symmetry. In fact, free fermion
systems are the simplest systems with non-trivial quantum order.  In this
section, we will study the quantum order in a free fermion system to gain some
intuitive understanding of quantum order.

To find quantum order, or to even define quantum order, we must find universal
properties.  The universal properties are the properties which do not change
under any perturbations of the Hamiltonian that do not affect the symmetry.
Once we find those universal properties, we can use them to group many-body
wave functions into universal classes such that the wave functions in each
class have the same universal properties. Hopefully those universal classes
correspond to quantum phases in a phase diagram.  To really show that those
universal classes do correspond to quantum phases, we must show that as we
deform the Hamiltonian to drive the ground state from one universality class
to another, the ground state energy always has a singularity at the transition
point. (For zero temperature quantum transition, the ground state energy play
the role of free energy for finite temperature transition.  A singularity in
the ground state energy signal a quantum phase transition.)

We know that symmetry is a universal property. The order determined by such a
universal property is our old friend -- the symmetry-breaking order.  So to
show the existence of new quantum order, we must find universal properties
that are different from symmetry.

Let us consider free fermion system with only the translation symmetry and the
$U(1)$ symmetry from the fermion number conservation.  The Hamiltonian has a
form
\begin{equation}
 H=\sum_{\<\v i\v j\>} \left(c^\dag_{\v i} t_{\v i\v j}c_{\v j} 
                             + h.c.\right)
\end{equation}
with $t_{\v i\v j}^*=t_{\v j\v i}$.  
The ground state is obtained by filling every negative energy level with one
fermion.  In general, the system contains several pieces of Fermi surfaces.

\begin{figure}
\centerline{
\includegraphics[width=3.2in]{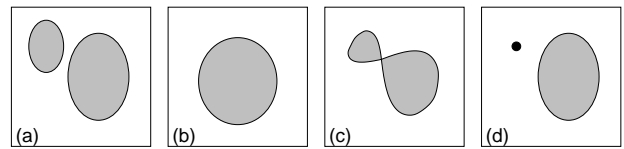}
}
\caption{
Two sets of oriented Fermi surfaces in (a) and (b) 
represent two different quantum orders. 
The two possible transition points between
the two quantum order (a) and (b) are
described by the Fermi surfaces (c) and (d).
}
\label{trans2d}
\end{figure}

We note that any small change of $t_{ij}$ do not change the topology of the
Fermi surfaces as long as the change do not break the translation symmetry and
do not violate the fermion number conservation.  So the the Fermi surface
topology is a universal properties.

To show that the Fermi surface topology really defines quantum phases, we need
to show that any change of the Fermi surface topology will lead to a
singularity in the ground state energy.  The Fermi surface topology can change
in two ways as we continuously changing $t_{ij}$: (a) a Fermi surface shrinks
to zero (Fig. \ref{trans2d}d) and (b) two Fermi surfaces join (Fig.
\ref{trans2d}c).  

When a Fermi surface
is about to disappear in a $D$-dimensional system, the ground state energy
density has a form
\begin{equation}
 \rho_E = \int \frac{d^D\v k}{(2\pi)^D} (\v k\cdot M \cdot \v k -\mu)
\Th( -\v k\cdot M \cdot \v k+\mu) + ...  \nonumber 
\end{equation}
where the $...$ represents non-singular contribution and the symmetric matrix
$M$ is positive (or negative) definite.  
The integral in the above equation simply represents the total energy
of the filled states enclosed by the small Fermi surface.
The small Fermi surface  is about to
shrink to zero as $\mu$ pass zero.
We find that the ground state energy density has a singularity at $\mu=0$:
$ \rho_E = c \mu^{(2+D)/2} \Th(\mu) + ...$,
where 
$ \Th(x>0) = 1$, $\Th(x<0) = 0$.
When two Fermi surfaces are about to join, the singularity is still
determined by the above equation, but now
$M$ has both negative and positive eigenvalues.
The ground state energy density has a singularity
$ \rho_E = c \mu^{(2+D)/2} \Th(\mu) + ...$ when $D$ is odd and
$ \rho_E = c \mu^{(2+D)/2} \log |\mu| + ...$ when $D$ is even.

We find that the ground state energy density has a singularity at $\mu=0$
which is exactly the same place where the topology of the Fermi surfaces has a
change \cite{L6030}.  Therefore the topology of the Fermi surface is a
universal property that define a order.  We note that the states with
different  Fermi surface topologies all have the same symmetry.  Thus the
quantum phase transition that change the the topology of the Fermi surface
does not change any symmetry.  Therefore the order defined by the Fermi
surface topology is a new kind of order that cannot be characterized by
symmetries. 
Such an order is an example of quantum order.

\section{ Quantum order in boson/spin liquids}
\label{prjcon}

\subsection{Quantum order and new universal properties}
\label{qouni}

After realizing the existence of the quantum order in free fermion systems,
we may expect quantum order to be a general phenomena.  In this section we
would like to study the existence of quantum order in interacting boson or
spin systems.  Instead of looking for universal properties, we would like
to first look for boson/spin states that contain emergent gauge bosons and
fermions.  The emergence of gauge bosons and fermions indicate the appearance
of new quantum order. Then we will study the universal properties which give a
more systematic description of quantum orders.

We like to point out that a spin system is a special case of boson system
since we can regard a site with a down spin as an empty site and a site with a
up spin as a occupied site for bosons.  In this section we will
interchangeably use both the boson and the spin languages to describe the same
system.

\subsection{Projective construction}

In the introduction, we argue that the existence of light and electron implies
that our vacuum contains a non-trivial quantum order.  However, we do not know
to which system does the quantum order belong.  Now we look for quantum order
in spin systems. So we know our system.  But we do not know what to look for,
since we have no clue what does a quantum order look like at microscopic level.
So instead of directly searching for quantum order, let us look for something
slightly more familiar: a spin liquid state that does not break any symmetry.  

\subsubsection{A mean-field theory of spin liquids}

To be concrete, let us consider a spin-1/2 system on a square lattice
\begin{equation}
\label{Hspin}
  H=\sum_{\<\v i\v j\>} J_{\v i\v j} \v S_{\v i} \cdot \v S_{\v j} .
\end{equation}
In the conventional mean-field theory, we use the ground 
$|\Phi^{\v m_{\v i}}_{mean}\>$ of a
free spin Hamiltonian
\begin{equation*}
 H_{mean}=\sum_{\<\v i\>} \v m_{\v i}\v S_{\v i}
\end{equation*}
to approximate the ground state of the interacting Hamiltonian $H$.  The
mean-field ground state described by $\v m_{\v i}=\bar{\v m_{\v i}}$,
$|\Phi^{\bar{\v m}_{\v i}}_{mean}\>$, is obtained by
minimizing the average energy $\<\Phi^{{\v m}_{\v i}}_{mean}|H|\Phi^{{\v
m}_{\v i}}_{mean}\>$.  However, no matter how we choose the mean-field ansatz
$\v m_{\v i}$, the mean-field ground state always break spin rotation symmetry
and there is no way to obtain a spin liquid.

We have to use another approach to obtain a spin liquid \cite{BZA8773,AM8874}.
We start with a free fermion mean-field Hamiltonian that contains
two fermion fields $\psi_{\v i}=\bpm \psi_{1\v i}\\ \psi_{2\v i} \epm$:
\begin{equation}
\label{Hmean}
 H_{mean}=\sum \psi_{\v i}^\dag u_{\v i\v j} \psi_{\v i}
\end{equation}
where $u_{\v i\v j}$ are two by two complex matrices defined on the links
$\<\v i\v j\>$ that describe the hopping of the fermions.  However, the
mean-field ground state of $H_{mean}$, $|\Psi^{u_{\v i\v j}}_{mean}\>$, does
not correspond to a spin state.  But, we can obtain a spin state from the
fermion state by projecting the fermion state $|\Psi^{u_{\v i\v j}}_{mean}\>$
into the subspace where every site has even numbers of fermion:
\begin{equation*}
 |\Phi_{spin}^{u_{\v i\v j}}\>=\cP |\Psi^{u_{\v i\v j}}_{mean}\>
\end{equation*}
This is because there are only two states, on each site, that have even number
of fermion. One is the empty site $|0\>$ which can be viewed as a spin-down 
state, and the other is the state with two fermions $\psi^\dag_{1\v
i}\psi^\dag_{2\v i}|0\>$  which corresponds to the spin-up state.

Although it is not obvious, one can show \cite{Wqoslpub} that iff
$u_{\v i\v j}$ satisfy
\begin{equation*}
 \Tr (u_{\v i \v j}) =\text{imaginary},\ \ \ \ \ \
 \Tr (u_{\v i \v j}\tau^l) =\text{real},\ \ \ \ l=1,2,3
\end{equation*}
where $\tau^{1,2,3}$ are the Pauli matrices, then $\Phi_{spin}^{u_{\v i\v j}}$
describes a spin rotation invariant state.  Since the spin state is obtained
through the projection $\cP$, we will call the above construction projective
construction.  It is a special case of the slave-boson construction
\cite{BZA8773,AM8874} at zero doping. We see that at least we can use the
projective construction to construct a spin liquid state which does not break
the spin rotation symmetry.

Through the projective construction, we introduced a label $u_{\v i\v j}$ that
labels a class of spin wave functions. ($u_{\v i\v j}$ does not label
all possible spin wave functions.)  So we do not have to directly deal with
the many-body functions of of spin liquids which are very hard to visualize.
We only need to deal with $u_{\v i\v j}$ to understand the properties of the
spin liquids.

\subsubsection{The variational ground state}

We may view $u_{\v i\v j}$ as variational parameters.  An approximate many-body
ground state wave function $|\Phi_{spin}^{\bar u_{\v i\v j}}\>$ can be
obtained by minimizing the average energy $\<\Phi_{spin}^{u_{\v i\v
j}}|H|\Phi_{spin}^{u_{\v i\v j}}\>$ where $H$ is the spin Hamiltonian.

Aside from its many variational parameters, there is no reason to expect the
projective construction to give a good approximation of the ground state for a
generic spin Hamiltonian. So there is no reason to trust any results
obtained from projective construction.  As we will see below that the
projective construction leads to many unbelievable results, such as the
emergence of gauge bosons, fermions, or even anyons from purely bosonic
systems. So those amazing results may just be the artifacts of a unreliable
approach.

However, for certain type of Hamiltonians, the projective construction leads
to a good approximation of the ground state.  In certain large $N$ limits
\cite{AM8874,RS9173}, the fluctuations around the mean-field ansatz are weak,
and the projective construction gives a good description of the ground state
and the excitations.  We can also construct special Hamiltonians where the
projective construction leads to an exact ground state (and all the exact
excited states. See section \ref{exct}).  For those Hamiltonians, the
projective construction does provide a good description of spin liquid states
which cannot be provided by other conventional method. In the following, we
will only consider those friendly Hamiltonians and trust the results of the
projective construction.

I would like to mention that in practice, most of the unbelievable predictions
from the projective construction turn out to be correct.  For example, in the
research of high $T_c$ superconductors, both the $d$-wave superconducting
state and the pseudo-gap metallic state was predicted by the projective
construction prior to experimental observation \cite{KL8842,AM8874}.  This may
be the first time in the history of condensed matter physics that a truly new
state of matter -- the pseudo-gap metallic state -- is predicted before the
experimental observation \cite{MDL9641}.

\subsubsection{Low energy excitations}

In the conventional mean-field theory for spin ordered state, after we obtain
the mean-field ground state $|\Phi^{\bar{\v m}_{\v i}}_{mean}\>$ that minimize
the average energy, we can create collective excitations above the ground
state through the fluctuation of the mean-field ansatz $\v m_{\v i} = \bar{\v
m}_{\v i}+\del {\v m}_{\v i}$.  Those collective excitations correspond to the
spin wave excitations.

In the projective construction, we can create collective excitations in the
exactly the same way.  The collective excitations above the mean-field ground
state $\Phi_{spin}^{\bar u_{\v i\v j}}$ correspond to the fluctuations of
the mean-field ansatz
$u_{\v i\v j}=\bar u_{\v i\v j}+\del u_{\v i\v j}$.  The physical spin
wave function for such type of excitations is obtained via the projection of
the deformed fermion state $|\Psi^{\bar u_{\v i\v j}+\del u_{\v i\v
j}}_{mean}\>$:
\begin{equation*}
 |\Phi_{spin}^{\del u_{\v i\v j}}\> =
\cP|\Psi^{\bar u_{\v i\v j}+\del u_{\v i\v j}}_{mean}\>
\end{equation*}

The ground state obtained from the projective construction also contains a
second type of excitations.  This type of excitations corresponds to 
fermion pair excitations.  We start with the fermion ground state with a pair
of particle-hole excitations $\psi^\dag_{a\v i}\psi_{b\v j}|\Psi^{\bar u_{\v
i\v j}}_{mean}\>$.  After the projection, we obtain the corresponding physical
spin state
\begin{equation*}
|\Phi_{spin}^{(a,\v i,b,\v j)}\>=
\psi^\dag_{a\v i}\psi_{b\v j}|\Psi^{\bar u_{\v i\v j}}_{mean}\>
\end{equation*}
that describes a pair of fermions.

Clearly the fermions excitations interact with the collective modes $\del
u_{\v i\v j}$. The effective Lagrangian that describes the two types of
excitations has a form $\cL(\psi, \del u_{\v i\v j})$.  It appears that the
spin liquid state obtained through the projective construction always contain
fermionic excitations described by $\psi$.  The emergent fermions will imply
that the spin liquid state is a new state of matter and contain non-trivial
quantum order.

However, the thing is not that easy. It turns out the collective fluctuations
represent gauge fluctuations and the fermions carry gauge charges.  Those
fluctuations can mediate an confining interactions between the fermions.  As a
result, the spin liquid state may not contain any fermionic excitations and
may not represent new state of matter. 

To see that the fluctuations of $u_{\v i\v j}$ represent gauge fluctuations,
we note that the mean-field Hamiltonian $H_{mean}$ is invariant under the
following $SU(2)$ gauge transformation
\begin{align}
\label{gaugetran}
\psi_{\v i} \to & W_{\v i}\psi_{\v i},\ \ \ \ \ W_{\v i}\in SU(2)  \nonumber \\
u_{\v i \v j} \to &W_{\v i}~u_{\v i \v j}~W^\dag_{\v j}    
\end{align}
So the two fermion ground states of the $H_{mean}$ corresponding to two ansatz
$u_{\v i\v j}$ and $u'_{\v i\v j}$
are related by an $SU(2)$ gauge transformation if $u'_{\v i\v j}=W_{\v
i}u_{\v i \v j}W^\dag_{\v j}$.  Since the even-fermion states $|0\>$ and
$\psi^\dag_{1\v i}\psi^\dag_{2\v i}|0\>$ are invariant under $SU(2)$ gauge
transformation, the projected state $|\Phi_{spin}^{u_{\v i\v j}}\>=\cP
|\Psi_{mean}^{u_{\v i\v j}}\> $ is invariant under the $SU(2)$ gauge
transformation:
\begin{equation*}
 |\Phi_{spin}^{u_{\v i\v j}}\>=
 |\Phi_{spin}^{W_{\v i}u_{\v i \v j}W^\dag_{\v j}}\>
\end{equation*}
As a result, $u_{\v i\v j}$ is not a one-to-one label of the physical spin
state, but a many-to-one label.  The mean-field energy $E(u_{\v i\v j})=
\<\Phi_{spin}^{u_{\v i\v j}}|H|\Phi_{spin}^{u_{\v i\v j}}\>$, as a function of
real physical spin state $|\Phi_{spin}^{u_{\v i\v j}}\>$, is invariant under
the $SU(2)$ gauge transformation $E(u_{\v i\v j})= E(W_{\v i}u_{\v i\v
j}W^\dag_{\v j})$.  Similarly, the effective Lagrangian $\cL(\psi, u_{\v i\v
j})$ is also invariant under the $SU(2)$ gauge transformation:
\begin{equation*}
 \cL(\psi, u_{\v i\v j})=\cL(W_{\v i}\psi, W_{\v i}u_{\v i\v j}W^\dag_{\v j})
\end{equation*}
The $SU(2)$ gauge invariance of the effective Lagrangian strongly affect the
dynamics of $u_{\v i\v j}$ fluctuations. It makes the $u_{\v i\v j}$
fluctuations to behave like $SU(2)$ gauge fluctuations.  If we write $u_{\v
i\v j}=i\chi e^{ia_{\v i\v j}^l\tau^l}$, then $a^l_{\v i\v j}$ play the role
of the $SU(2)$ gauge potential on the lattice.

\subsection{Deconfined phase and new state of matter}

We have mentioned that to obtain a new state of matter from the projective
construction, the gauge fluctuations $u_{\v i\v j}$ must not mediate a
confining interaction.  One may say that the $SU(2)$ gauge fluctuations always
mediate a confining interaction in 1+2D, so the projective construction can
never produce a spin liquid with emergent fermions.  Well again thing is not
that simple.  It turns out that whether the gauge fluctuations confine the
fermions or not depend on the form of the mean-field ansatz $\bar u_{\v i\v
j}$ that minimize the average energy $\<H\>$. 

To understand how the ansatz $\bar u_{\v i\v j}$ affect the
dynamics of the gauge fluctuations, it is convenient to introduce the loop
variable
\begin{equation}
\label{su2flux}
P(C_{\v i})= \bar u_{\v i\v j}\bar u_{\v j\v k}...  \bar u_{\v l\v i}
\end{equation}
If we write $P(C_{\v i})$ as $P(C_{\v i})=\chi e^{i\Phi^l(C_{\v i})\tau^l}$,
then $\Phi^l\tau^l$ is the $SU(2)$ flux through the loop $C_{\v i}$: $\v i\to
\v j\to \v k\to ..\to \v l\to \v i$ with base point $\v i$.
The $SU(2)$ flux correspond to the gauge field strength in the continuum
limit.

\begin{figure}
\centerline{
\includegraphics[width=2in]{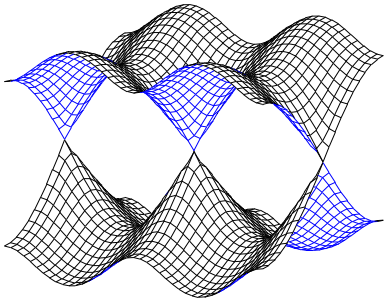}
}
\caption{
The fermion dispersion for the ansatz \Eq{piF}.
}
\label{piFdis}
\end{figure}

\subsubsection{$SU(2)$ spin liquid}

If for a certain spin Hamiltonian $H$, $\bar u_{\v i\v j}$ has a form
\begin{align}
\label{piF}
u_{\v i, \v i + \v x} & = - i(-)^{i_y} \chi ,
\nonumber\\
u_{\v i, \v i + \v y} & = - i\chi ,
\end{align}
the $SU(2)$ flux through any loops is trivial. In this case the fluctuations
of $u_{\v i\v j}$ behave like the usual $SU(2)$ gauge fluctuations. The
fermion dispersion is determined by mean-field Hamiltonian $H_{mean}$ in
\Eq{Hmean}: $E_{\v k}=\pm 2\chi \sqrt{\sin^2 k_x + \sin^2 k_y}$.  At low
energies ($E_{\v k}\sim 0$) the fermions have a linear dispersion $E \propto
|\v k|$ (see Fig. \ref{piFdis}) and can be described by massless Dirac
fermions in the continuum limit.  So in the continuum limit, the low energy
fluctuations are described by
\begin{equation}
\label{QCD3}
 \cL = \sum_{\sigma=1}^N\bar{\psi}_{\sigma}
(\partial_{\mu}-ia^l_{\mu}\tau^l)\gamma_{\mu}\psi_{\sigma}
+\frac{1}{2g} \Tr f^l_{\mu\nu}f^{l,\mu\nu}
\end{equation}
where $a^l_\mu$ and $f^l_{\mu\nu}$, $l=1,2,3$ the vector potential and the
field strength of the $SU(2)$ gauge field.

It is interesting to see that for the spin liquid described by the ansatz
\Eq{piF}, the low energy fluctuations is described by 1+2D QCD!  The low
energy physical properties of 1+2D QCD is complicated. 

But the above
construction can be easily generalize to higher dimensions.  The low energy
properties of 1+4D QCD is much easier to determine and it contains deconfined
phase. As a result, the 1+4D spin liquid contains emergent massless fermions
and emergent massless $SU(2)$ gauge bosons.  Such a state represent a new
state of matter and contains a non-trivial quantum order.  We get what we are
looking for (if we live in four dimensional space).  Because of the low energy
$SU(2)$ gauge fluctuations, we will call the ansatz \Eq{piF} $SU(2)$ ansatz
and the correspond spin liquid $SU(2)$ spin liquid.

\subsubsection{$U(1)$ spin liquid}

If the ansatz that minimizes the average energy has a form
\begin{align}
\label{sF}
u_{\v i, \v i + \v x} & = - \tau^3 \chi - i (-)^{\v i} \Delta ,
\nonumber\\
u_{\v i, \v i + \v y} & = - \tau^3 \chi + i (-)^{\v i} \Delta ,
\end{align}
where $(-)^{\v i}\equiv (-)^{i_x+i_y}$, then the low energy $u_{\v i\v j}$
fluctuations are actually described by $U(1)$ gauge fluctuations. This is
because the above ansatz contains non-trial $SU(2)$ flux: $P(C_{\v i})\propto
e^{i\Phi^3\tau^3}$. Unlike the flux of $U(1)$ gauge field, the flux of $SU(2)$
gauge field is not invariant under the $SU(2)$ gauge transformations. Instead,
the $SU(2)$ flux transforms like a Higgs field that carries non-zero $SU(2)$
charge. The non-zero $SU(2)$ flux has a similar effect as the condensation of
a Higgs field, which can give gauge bosons a mass term via the Anderson-Higgs
mechanism \cite{A6339,H6432}.  For our case, the $SU(2)$ flux in the $\tau^3$
direction give the $a^1_\mu$ and $a^2_\mu$ components of the gauge field a
mass, but $a^3_\mu$ remains massless. So the $SU(2)$ flux break the $SU(2)$
gauge structure down to a $U(1)$ gauge structure \cite{Wsrvb}. This is why the
spin liquid described by the ansatz \Eq{sF} contains only massless $U(1)$
gauge fluctuations.  The low energy fermions are still described by massless
Dirac fermions.  So the low energy effective theory of the spin liquid is a
1+2D QED
\begin{equation}
\label{QED3}
 \cL = \sum_{\sigma=1}^N\bar{\psi}_{\sigma}
(\partial_{\mu}-ia^3_{\mu}\tau^3)\gamma_{\mu}\psi_{\sigma}
+\frac{1}{2g} f^3_{\mu\nu}f^{3,\mu\nu}
\end{equation}
where $a^3_\mu$ and $f^3_{\mu\nu}$ are the vector potential and the field
strength of the $U(1)$ gauge field.  Again, the above construction can be
easily generalize to higher dimensions.  Now we do not need to go to 1+4
dimensions.  In 1+3 dimensions,  the spin liquid \cite{Wlight} already contains
emergent massless fermions and emergent massless $U(1)$ gauge bosons since
1+3D QED is not confining.  Such a 1+3D spin liquid represents another new
state of matter and  will be called $U(1)$ spin liquid. 

The close resemblance of the low energy effective theory of the spin liquids
and the QED/QCD in the standard model makes one really wonder: is this how
the QED and QCD emerge to be the effective theory that describe our vacuum
\cite{Wqoem}?

Even in 1+2D, the $U(1)$ spin liquid was shown to be a stable phase
\cite{Wqoslpub,RWspin,HSF0451} based a combined analysis of instanton
\cite{IL8988} and projective symmetry \cite{Wqoslpub} (see section \ref{psg}).
The existence of the $U(1)$ spin liquid is a striking phenomenon since the
gapless excitations interact down to zero energy, and yet remain to be
gapless.  The interaction is so strong that that are no free fermionic or
bosonic quasiparticles ar low energies. Since the $U(1)$ gauge bosons and the
fermions are not well defined at any energy, the $U(1)$ spin liquid was more
correctly called the algebraic spin liquid \cite{Wqoslpub,RWspin}.

Since there is no spontaneous broken symmetry to protect the above interacting
gapless excitations,  there should be a ``principle'' that prevents the
gapless excitations from opening an energy gap and makes the algebraic spin
liquids stable.  \Ref{Wqoslpub} proposed that quantum order is such a
principle. To support this idea, it was shown that just like the symmetry
group of symmetry breaking order protects gapless Nambu-Goldstone modes, the
projective symmetry group (see section \ref{psg}) of quantum order protects the
interacting gapless excitations in the algebraic spin liquid.  This result
implies that the stabilities of algebraic spin liquids are protected by their
projective symmetry groups.  The existence of gapless excitations without
symmetry breaking is a truly remarkable feature of quantum ordered states.  

\subsubsection{$Z_2$ spin liquid}

For the ansatz \cite{Wsrvb}
\begin{align}
\label{Z2gA}
u_{\v i,\v i+\v x }=&  u_{ \v i,\v i+\v y }= -\chi \tau^3 , \nonumber \\
u_{\v i,\v i+ \v x + \v y} =&  \eta\tau^1 +\la\tau^2 ,\nonumber \\
u_{\v i,\v i-\v x + \v y} =&  \eta\tau^1 -\la\tau^2 ,
\end{align}
the $SU(2)$ flux $\Phi^l(C_{\v i})\tau^l$ for different loops points in
different directions in the $( \tau^1, \tau^2, \tau^3)$ space.  The
non-collinear $SU(2)$ flux break the $SU(2)$ gauge structure down to a $Z_2$
gauge structure. Since the $Z_2$ gauge fluctuations only mediate a short
ranged interaction, the fermions are not confined even in 1+2D. The spin
liquid obtained from the ansatz \Eq{Z2gA} is a new state of matter that has
emergent fermions and $Z_2$ gauge theory.  Non-trivial quantum order can 
appear in two dimensional space.  The spin liquid will be called $Z_2$ spin
liquid. Such a spin liquid corresponds to the short-ranged Resonating Valence
Bound state proposed in \Ref{KRS8765,RK8876}.

\subsubsection{Summary}

The projective construction is a powerful way construct states
that represent new state of matter. Those states have emergent fermions and
gauge bosons, and thus contain a new kind of order that cannot be described by
symmetry. The new order is called quantum orders.  Certainly, not all states
obtained via the projection construction contain non-trivial quantum orders.
But many of them do.  The projective construction not only can produces states
with emergent QED and QCD, it can also produce states with fractional
statistics (including non-Abelian statistics) \cite{WWZcsp,J9153,Wpcon}.

\subsection{Quantum order and projective symmetry group}
\label{psg}

We know that different symmetry-breaking orders can be systematically
characterize by different symmetry group. The group theory description allows
us to classify 230 different 3D crystals.  Knowing the existence of new
quantum order in spin liquids, we would like to ask what mathematical object
that we can use to systematically describe different quantum orders?  In this
subsection, we are going to introduce a mathematical object -- projective
symmetry group and show that the projective symmetry group can (partially)
characterize different quantum orders.

\subsubsection{
The difficulty of seeing quantum order
}

In subsection \ref{qouni}, we argue that to describe or to even define quantum
order, we must find universal properties that are different from the symmetry
(such as the topology of the Fermi surfaces discussed in the section
\ref{freeferm}).  However, it is very difficult to find new universal
properties of generic many-body wave functions.  Let us consider the free
fermion systems that we discussed before to gain some intuitive understanding
of the difficulty.  We know that a free fermion ground state is described by
an anti-symmetric wave function of $N$ variables. The anti-symmetric function
has a form of Slater determinant: $\Psi(x_1,...,x_N)=\det(M)$ where the matrix
elements of $M$ is given by $M_{mn} =\psi_n(x_m)$ and $\psi_n$ are
single-fermion wave functions. The first step to find quantum orders in  free
fermion systems is to find a reasonable way to group the Slater-determinant
wave functions into classes.  This is very difficult to do if we only know the
real space many-body function $\Psi(x_1,...,x_N)$. However, if we use Fourier
transformation to transform the real-space wave function to momentum-space
wave function, then we can group different wave functions into classes
according to their Fermi surface topologies.  This leads to our understanding
of quantum orders in free fermions systems (see section \ref{freeferm}). The
Fermi surface topology is the quantum number that allows us to characterize
different quantum phases of free fermions.  Here we would like to stress that
without the Fourier transformation, it would be very difficult to see Fermi
surface topologies from the real space many-body function $\Psi(x_1,...,x_N)$.

For the boson/spin systems, what is missing here is the corresponding
``Fourier'' transformation.  Just like the topology of Fermi surface, it is
very difficult to see universal properties (if any) directly from the real
space wave function.  At moment there are two ways to understand the quantum
order in boson/spin systems. The first one is through the projective symmetry
group which will be discussed below.  The second one is through string-net
condensation which will be discuss in section \ref{str}.  Both the
projective symmetry group and the string-net condensation play the role of the
Fourier transformation in the free fermion system. They allow us the extract
the universal properties from the very complicated many-body wave functions.

\subsubsection{Symmetry of the spin liquids}

To motivate the projective symmetry group, let us first consider the symmetry
of the spin liquid states obtained from the $SU(2)$, $U(1)$ and $Z_2$ ansatz
\Eq{piF}, \Eq{sF}, and \Eq{Z2gA}.  At first sight, those spin liquids appear
not to have all the symmetries.  For example, the $U(1)$ ansatz \Eq{sF} is not
invariant under the translation in the $x$-direction. 

However, those ansatz do describe spin states that have all the
symmetries of square lattice, namely the two translation symmetries $T_x$:
$(i_x,i_y)\to (i_x+1,i_y)$ and $T_y$: $(i_x,i_y)\to (i_x,i_y+1)$, and three
parity symmetries, $P_{x}$: $(i_x, i_y)\to (-i_x,i_y)$, $P_{y}$: $(i_x,
i_y)\to (i_x,-i_y)$, and  $P_{xy}$: $(i_x, i_y)\to (i_y,i_x)$.  This is
because the ansatz $u_{\v i\v j}$ is a many-to-one label of the physical spin
state. The non-invariance of the ansatz does not imply the non-invariance of
the corresponding physical spin state after the projection.  We only require
the mean-field ansatz to be invariant up to a $SU(2)$ gauge transformation in
order for the projected physical spin state to have a symmetry. For example, a
$T_x$  translation transformation changes the $U(1)$ ansatz \Eq{sF} to
\begin{align*}
U_{\v i, \v i + \v x} & = - \tau^3 \chi + i (-)^{\v i} \Delta ,
\nonumber\\
U_{\v i, \v i + \v y} & = - \tau^3 \chi - i (-)^{\v i} \Delta ,
\end{align*}
The translated ansatz can be transformed into the original ansatz via a
$SU(2)$ gauge transformation $W_{\v i}=(-)^{\v i}i\tau^1$.  Therefore, after
the projection, the ansatz \Eq{sF} describes a $T_x$ translation symmetric
spin state.  

Using the similar consideration, one can show that the $SU(2)$, $U(1)$, and
$Z_2$ ansatz are invariant under translation $T_{x,y}$ and parity $P_{x,y,xy}$
symmetry transformations followed by corresponding $SU(2)$ gauge
transformations $G_{T_x,T_y}$ and $G_{P_x,P_y,P_{xy}}$ respectively. Thus the
three ansatz all describe symmetric spin liquids.  In the following, we list
the corresponding gauge transformations $G_{T_x,T_y}$ and $G_{P_x,P_y,P_{xy}}$
for the three ansatz:\\ 
for the $SU(2)$ ansatz \Eq{piF}:
\begin{align}
\label{GspiF}
G_{T_x}(\v i) =& (-)^{i_x}G_{T_y}(\v i) = \tau^0,  
&
G_{P_{xy}}(\v i) =&  (-)^{i_xi_y} \tau^0, 
\nonumber\\
(-)^{i_x}G_{P_x}(\v i) =& (-)^{i_y}G_{P_y}(\v i) = \tau^0, 
& 
G_0(\v i) =& e^{i\th^l\tau^l}
\end{align}
for the $U(1)$ ansatz \Eq{sF}:
\begin{align}
\label{GssF}
G_{T_x}(\v i) =& G_{T_y}(\v i) = i(-)^{\v i}\tau^1,  
&
G_{P_{xy}}(\v i) =&  i(-)^{\v i}\tau^1, 
\nonumber\\
G_{P_x}(\v i) =& G_{P_y}(\v i) = \tau^0, 
& 
G_0(\v i) =& e^{i\th\tau^3}
\end{align}
for the $Z_2$ ansatz \Eq{Z2gA}:
\begin{align}
\label{GsZ2gA}
G_{T_x}(\v i) =& G_{T_y}(\v i) = i\tau^0,  
&
G_{P_{xy}}(\v i) =&  \tau^0, 
\nonumber\\
G_{P_x}(\v i) =& G_{P_y}(\v i) = (-)^{\v i}\tau^1, 
& 
G_0(\v i) =& -\tau^0
\end{align}
In the above we also list the pure gauge transformation $G_0(\v i)$ that leave
the ansatz invariant: $\bar u_{\v i\v j}=G_0(\v i)\bar u_{\v i\v j}G^\dag_0(\v
j)$.

\subsubsection{Definition of PSG}

The $SU(2)$, $U(1)$ and $Z_2$ ansatz 
after the projection, give
rise to three spin liquid states. The three states have the exactly the same
symmetry. The question here is whether the three spin liquids belong to
the same phase or not.  According to Landau's symmetry breaking theory, two
states with the same symmetry belong to the same phase.  However, we now know
that Landau's symmetry breaking theory does not describe all the phases.  It
is possible that the three spin liquids contain different orders that cannot
be characterized by symmetries. The issue here is to find a new set of quantum
numbers that characterize the new orders. 

To find a new set of universal quantum numbers that distinguish the three spin
liquids, we note that although the three spin liquids have the same symmetry,
their ansatz are invariant under the symmetry translations followed by
different gauge transformations (see \Eq{GspiF}, \Eq{GssF}, and \Eq{GsZ2gA}).
So the invariant group of three ansatz are different.  We can use the
invariant group of the three ansatz to characterize the new order in the spin
liquid. In a sense, the invariant group define a new order --
quantum order.

The invariant group of an ansatz is formed by all the combined symmetry
transformations and the gauge transformations that leave the ansatz invariant.
Those combined transformations from a group. Such a group is called the
Projective Symmetry Group (PSG).  The combined transformations $(G_{T_x}T_x,
G_{T_y}T_y, G_{P_x}P_x, G_{P_y}P_y, G_{P_{xy}}P_{xy}) $ and $G_0$ in
\Eq{GspiF}, \Eq{GssF}, and \Eq{GsZ2gA} generate the three PSG's for the three
ansatz \Eq{piF}, \Eq{sF}, and \Eq{Z2gA}.

\subsubsection{Properties of PSG}

To understand the properties of the PSG, we would like to point out that a PSG
contains a special subgroup, which will be called the invariant gauge group
(IGG). An IGG is formed by pure gauge transformations that leave the ansatz
unchanged \begin{align} IGG\equiv \{ G_0 |\ & u_{\v i\v j}= G_0(\v i) u_{\v
i\v j} G_0^\dag(\v j) \} \nonumber \end{align} For the ansatz \Eq{piF},
\Eq{sF}, and \Eq{Z2gA}, the IGG's are $SU(2)$, $U(1)$, and $Z_2$ respectively.
We note that $SU(2)$, $U(1)$, and $Z_2$ happen to be the gauge groups that
describe the low energy gauge fluctuations in the three spin liquids
This relation is not an accident.  In general the gauge group of the low
energy gauge fluctuations for a spin liquid described by an ansatz $\bar u_{\v
i\v j}$ is given by the IGG of the ansatz \cite{Wqoslpub}. This result
generalizes the analysis of the low energy gauge group based on the $SU(2)$
flux.

If an ansatz is invariant under the translation $T_x$ followed by a gauge
transformation $G_x$, then it is also invariant under the translation $T_x$
followed by another gauge transformation $G_0G_x$, as long as $G_0 \in IGG$. So the
gauge transformation associated with a symmetry transformation is not unique.
The number of the choices of the gauge transformations is the number of the
elements in IGG.  We see that as sets, $PSG=SG\times IGG$ where SG is the
symmetry group.  But as groups, $PSG$ is not the direct product of $SG$ and
$IGG$.  It is a ``twisted''  product. Using the more rigorous mathematical
notation, we have
\begin{equation*}
 SG=PSG/IGG
\end{equation*}
We may also say that the $PSG$ is a projective extension of $SG$ by $IGG$.

The $SU(2)$, $U(1)$ and $Z_2$ ansatz 
all have the same symmetry and hence the same symmetry group $SG$. They have
different $PSG$'s since the same $SG$ is extended by different $IGG$'s.  Here
we would like to remark that even for a given pair of $SG$ and $IGG$, there
are many different ways to extent the $SG$ by the $IGG$, leading to many
different $PSG$'s.  For example there are over 100 ways to extend the symmetry
group of a square lattice by a $Z_2$ IGG. This implies that there are over 100
different $Z_2$ spin liquids on a square lattice and those spin liquids all
have the exactly the same symmetry!  Finding different ways of extending a
symmetry group $SG$ is a pure mathematical problem. Such a calculation will
lead to a (partial) classification of the quantum orders (and the spin
liquids).

\subsubsection{PSG is a universal property which protects gapless excitations}

From the above discussion, we see that a PSG 
contains two parts. The first part is SG which describe the symmetry of the
spin liquid. The second part is IGG which describe the gauge ``symmetry'' of
the spin liquid. A generic elements in the PSG is a combination of the
symmetry transformation and the gauge transformation.

We know that symmetry and gauge ``symmetry'' are universal properties, \ie
perturbative fluctuations cannot break the symmetry, nor can they break the
gauge ``symmetry''. So both SG and IGG are universal properties.  This
strongly suggests that the PSG is also a universal property.

To directly show a PSG to be a universal property, we note that the fermion
mean-field Hamiltonian $H_{mean}$ in \Eq{Hmean} is invariant under the lattice
symmetry and the $SU(2)$ gauge transformations \Eq{gaugetran}.  But the
mean-field ansatz $\bar u_{\v i\v j}$ is not invariant under the separate
lattice symmetry and $SU(2)$ gauge transformations.  So the mean-field state
break the separate lattice symmetry and $SU(2)$ gauge ``symmetry'' down to a
smaller symmetry. The symmetry group of this smaller symmetry is the PSG.  So,
the PSG is the symmetry of the mean-field theory with $\bar u_{\v i\v j}$
ansatz.
As a result, the PSG is the symmetry for the effective Lagrangian $\cL_{\bar
u_{\v i\v j}}(\psi,\del u_{\v i\v j})$ that describes the fluctuations around
the mean-field ansatz.  If the mean-field fluctuations do not have any
infrared divergence, then those fluctuations will be perturbative in nature
and cannot change the symmetry -- the PSG. 

What do we mean by ``perturbative fluctuations cannot change the PSG?'' We
know that a mean-field ground state is characterized by $\bar u_{\v i\v j}$.
If we include perturbative fluctuations to improve our calculation of the
mean-field energy $\<\Phi_{spin}^{u_{\v i\v j}}|H|\Phi_{spin}^{u_{\v i\v
j}}\>$, then we expect the $\bar u_{\v i\v j}$ that minimize the improved
mean-field energy to receive perturbative corrections $\del \bar u_{\v i\v
j}$.  The statement ``perturbative fluctuations cannot change the PSG?'' means
that $\bar u_{\v i\v j}$ and $\bar u_{\v i\v j}+\del \bar u_{\v i\v j}$ have
the same PSG.

As the perturbative fluctuations (by definition) do not change the phase,
$\bar u_{\v i\v j}$ and $\bar u_{\v i\v j}+\del \bar u_{\v i\v j}$ describe
the same phase.  In other words, we can group $\bar u_{\v i\v j}$ into classes
(which are called universality classes) such that the $\bar u_{\v i\v j}$ in
each class are connected by the perturbative fluctuations.  
By definition, each universality
class describes one phase.  We see that, if the above argument about the
universality of the PSG's is true, then
the ansatz in a universality class all share the same PSG. In other words, the
universality classes (or the phases) are classified by the PSG's.  Thus the
PSG is a universal property.  We can use the PSG to describe the quantum order
in the spin liquid, as long as the low energy effective theory $\cL_{\bar
u_{\v i\v j}}(\psi, \del u_{\v i\v j})$ does not have any infrared divergence.  

In the standard renormalization group analysis of the stability of a phase or
a critical point, one needs to include all the counter terms that have the
right symmetries into the effective Lagrangian, since those terms can be
generated by perturbative fluctuations.  Then we examine if those allowed
counter terms are relevant perturbations or not.  In our problem, $\del \bar
u_{\v i\v j}$ discussed above correspond to thoe counter terms. The  effective
Lagrangian with the counter term is given by $\cL_{\bar u_{\v i\v j}+\del\bar
u_{\v i\v j}}(\psi, \del u_{\v i\v j})$.  The new feature here is that it is
incorrect to use the symmetry group alone to determine the allowed counter
terms $\del \bar u_{\v i\v j}$.  We should use PSG to determine the allowed
counter terms. In our analysis of the stability of phases and critical points, 
only the counter terms  $\del \bar u_{\v i\v j}$ that do not change the PSG
of $\bar  u_{\v i\v j}$ are allowed.

The $Z_2$ spin liquid \Eq{Z2gA} (and other 100 plus $Z_2$ spin liquids)
contains no diverging fluctuations. So the PSG description of the quantum
order is valid for this case.  For the $SU(2)$ spin liquid \Eq{piF} and the
$U(1)$ spin liquid \Eq{sF}, their low energy effective theory \Eq{QCD3} and
\Eq{QED3} contain $\log$ divergence.  These are marginal cases where the PSG
description of the quantum order still apply. In a renormalization group
analysis of the stability of the $U(1)$ spin liquid \Eq{sF}, one can show
that,  in a large $N$ limit, the counter terms allowed by the $U(1)$ PSG
\Eq{GssF} are all irrelevant \cite{RWspin,HSF0451}, even if we include the
instanton effect \cite{IL8988}.  Thus the (large $N$)  $U(1)$ spin liquid is a
stable quantum phase.  One can also show that non of the allowed counter terms
can give the gapless fermions and gapless gauge bosons an energy gap
\cite{Wqoslpub,WZqoind}.  Thus the gapless excitations in the  $U(1)$ spin
liquid are protected by by the $U(1)$ PSG, despite those gapless excitations
interact down to zero energy.

\subsection{An intuitive understand of quantum order and the emergent gauge
bosons and fermions}

The projective construction produces a correlated many-body ground state
$|\Phi_{spin}\> = \cP |\Psi_{mean}^{\bar u_{\v i\v j}}\>$. We may view the
complicated correlation in the ground state as a pattern of quantum
entanglement.  The quantum order and the associated PSG is a characterization
of such a pattern of entanglement.  The gauge fluctuation above the many-body
ground state can be viewed as a fluctuation of the entanglement.  The fermion
excitations can be viewed as topological defects in the entanglement.  From
this point of view, the theory of quantum order can be regarded as a theory of
many-body quantum entanglement.

\section{An exact soluble model from projective construction}
\label{exct}

Usually, the projective construction does not give us exact results. In this
section, we are going to construct an exactly soluble model on 2D square
lattice \cite{K032,Wqoexct}. The model has a property that the projective
construction give us exact ground states and all other exact excited states.

\subsection{An exact soluble model for the $\psi$-fermions}

First, we would like to construct an exact soluble model for the
$\psi$-fermions.  It is convenient to write the exact soluble Hamiltonian in
terms of four Majorana fermions
\begin{equation}
\label{psila}
2 \psi_{1,\v i} = \la^{x}_{\v i} +i\la^{\bar x}_{\v i}, \ \ \ \ \ \
2 \psi_{2,\v i} = \la^{y}_{\v i} +i\la^{\bar y}_{\v i}
\end{equation}
The Majorana fermions satisfy the algebra $\{ \la_{a,\v i},\la_{b,\v j} \} =
2\del_{ab}\del_{\v i\v j}$.  where $a,b = x,\bar x, y, \bar y$.
The exact soluble fermion Hamiltonian is given by
\begin{align}
\label{HgF}
 H =& -\sum_{\v i} g \hat F_{\v i}, \\
\hat F_{\v i} = &
\hat U_{\v i,\v i+\v x} 
\hat U_{\v i+\v x, \v i+\v x+\v y} 
\hat U_{\v i+\v x+\v y,\v i+\v y} 
\hat U_{\v i+\v y,\v i} ,  \nonumber\\
\hat U_{\v i,\v i+\hat{\v x}}=&\la^{x}_{\v i}\la^{\bar x}_{\v i+\hat{\v x}},
\ \ \ \
\hat U_{\v i,\v i+\hat{\v y}}=\la^{y}_{\v i}\la^{\bar y}_{\v i+\hat{\v y}},
\ \ \ \
\hat U_{\v i\v j} =& -\hat U_{\v j\v i}.
\nonumber 
\end{align}

To see why the above interacting fermion model is exactly soluble, we note
that $\hat U_{\v i\v j}$ commute with each other and $H$ commute with all the
$\hat U_{\v i\v j}$.  So we can find the eigenvalues and eigenstates of $H$ by
finding the common eigenstates of $\hat U_{\v i\v j}$:
\begin{equation*}
\hat U_{\v i\v j}
|\{s_{\v i\v j} \}\> 
=s_{\v i\v j} |\{s_{\v i\v j} \}\> 
\end{equation*}
Since $(\hat U_{\v i\v j})^2=-1$ and $\hat U_{\v i\v j}=-\hat U_{\v j\v i}$ ,
$s_{\v i\v j}$ satisfies $s_{\v i\v j}=\pm i$ and $s_{\v i\v j}=-s_{\v j\v
i}$.  Since $H$ is a function of $\hat U_{\v i\v j}$'s, $|\{s_{\v i\v j}\}\>$
is also an energy eigenstate of \Eq{HgF} with energy
\begin{align} 
\label{Eng}
E = & -\sum_{\v i} g F_{\v i}
,  \nonumber\\
F_{\v i} = &
s_{\v i,\v i+\v x} 
s_{\v i+\v x, \v i+\v x+\v y} 
s_{\v i+\v x+\v y,\v i+\v y} 
s_{\v i+\v y,\v i} .
\end{align} 

To see if $|\{s_{\v i\v j}\}\>$'s represent all the exact eigenstates of $H$,
we need count the states.  Let us assume the 2D square lattice to have
$N_{site}$ lattice sites and a periodic boundary condition in both directions.
In this case the lattice has $2N_{site}$ links.  Since there are total of
$2^{2N_{site}}$ different choices of $s_{\v i\v j}$ (two choices for each
link), the states $|\{s_{\v i\v j}\}\>$ exhaust all the $4^{N_{site}}$ states
in the $(\psi_1,\psi_2)$ Hilbert space.  Thus the common eigenstates of $\hat
U_{\v i\v j}$ is not degenerate and the above approach allows us to obtain all
the eigenstates and eigenvalues of the $H$. 

We note that the eigenstate $|\{s_{\v i\v j}\}\>$ is the ground state
of the following free fermion Hamiltonian
\begin{equation}
\label{Hsij}
 H_{mean}=\sum_{\<\v i\v j\>}  s_{\v i\v j} \hat U_{\v i\v j}
\end{equation}
by choosing different $s_{\v i\v j}=\pm i$, the ground state of the above
mean-field Hamiltonian give rise to all the eigenstates of the interacting
fermion Hamiltonian \Eq{HgF}.

\subsection{An exact soluble spin-1/2 model}

We note that the Hamiltonian $H$ can only change the fermion number on each
site, $n_{\v i}=\psi^\dag_{1,\v i} \psi_{1,\v i} +\psi^\dag_{2,\v i}
\psi_{2,\v i}$, by an even number. Thus the $H$ acts within a subspace which
has an even number of fermions on each site. We will call such a subspace
physical Hilbert space. The physical Hilbert space has only two states per
site corresponding to a spin-up and a spin-down state.  
When restricted within the physical space, $H$ actually describes a
spin-1/2 system.  To obtain the corresponding spin-1/2 Hamiltonian, we note
that 
\begin{align}
\label{taupsi}
\si^{x}_{\v i} = i\la^y_{\v i}\la^x_{\v i},\ \ \ \ \ \ 
\si^{y}_{\v i} = i\la^{\bar x}_{\v i}\la^y_{\v i},\ \ \ \ \ \ 
\si^{z}_{\v i} = i\la^x_{\v i} \la^{\bar x}_{\v i}
\end{align}
act within the physical Hilbert space and satisfy the algebra of Pauli
matrices.  Thus we can identify $\si^l_{\v i}$ as the spin operator.  Using
the fact that
\begin{equation*}
(-)^{ n_{\v i} }
=\la^x_{\v i} \la^y_{\v i} \la^{\bar x}_{\v i} \la^{\bar y}_{\v i}
=1 
\end{equation*}
within the physical Hilbert space, we can show that the fermion Hamiltonian
\Eq{HgF} becomes (see Fig. \ref{t3string})
\begin{equation}
\label{spinH}
 H_{spin}= - \sum_{\v i} g \hat F_{\v i}, \ \ \ \ \ \ \ \hat F_{\v i}=
\si^x_{\v i} \si^y_{\v i+\hat{\v x}}
\si^x_{\v i+\hat{\v x}+\hat{\v y}} \si^y_{\v i+\hat{\v y}}
\end{equation}
within the physical Hilbert space.

\subsection{The projective construction leads to exact results}

\index{physical state}
\index{physical Hilbert space}
All the states in the physical Hilbert space (\ie all the states in the
spin-1/2 model) can be obtained from the $|\{s_{\v i\v j}\}\>$
states by projecting into the physical Hilbert space:
$\cP|\{s_{\v i\v j}\}\>$. The projection operator is given by
\begin{equation*}
 \cP = \prod_{\v i} \frac{1+(-)^{n_{\v i}}}{2}
\end{equation*} 
Since $[\cP, H]=0$, the projected state $\cP|\{s_{\v i\v j}\}\>$, if non-zero,
remain to be an eigenstate of $H$ (or $H_{spin}$) and remain to have the same
eigenvalue.  We see that the ground state of the mean-field Hamiltonian
\Eq{Hsij}, after the projection, give rise to all the exact eigenstates of the
spin Hamiltonian \Eq{spinH}. The project construction is exact for \Eq{spinH}!

\subsection{The $Z_2$ gauge structure}

The physical states (with even numbers of fermions per site) are invariant
under local $Z_2$ transformations generated by
\begin{equation*}
 \hat G=\prod_{\v i} G_{\v i}^{n_{\v i}}
\end{equation*}
where $G_{\v i}$ is an arbitrary function with only two values $\pm 1$.  
The $Z_2$ transformation change $\psi_{a\v i}$ to $\t \psi_{a\v
i}=G_{\v i}\psi_{a\v i}$ and $s_{\v i\v j}$ to $\t s_{\v i\v j}=G_{\v i}s_{\v
i\v j} G_{\v j}$, or more precisely
\begin{equation*}
 |\{\t s_{\v i\v j}\}\> = \hat G |\{ s_{\v i\v j}\}\> 
\end{equation*}
Using $\cP \hat G=\cP$,
we find that $|\{s_{\v i\v j}\}\>$ and $|\{\t s_{\v i\v
j}\}\>$ give rise to the same physical state after projection (if their
projection is not zero):
\begin{equation}
\label{PsPts}
 \cP |\{\t s_{\v i\v j}\}\> = \cP |\{ s_{\v i\v j}\}\> 
\end{equation}
Thus, $s_{\v i\v j}$ is a many-to-one label of the physical spin state.  

The above results indicate that we can view $is_{\v i\v j}$ as a $Z_2$ gauge
potential and the local $Z_2$ transformation is a $Z_2$ gauge transformation.
\Eq{PsPts} implies that gauge equivalent gauge potential described the same
physical state.  The fluctuations of $s_{\v i\v j}$ is described by a $Z_2$
gauge theory, which is the low energy effective theory of the spin system
\Eq{spinH}.

\section{Closed-string condensation}
\label{str}

The ground state of the exactly soluble model contains a special property --
closed-string condensation. In this section, we will see that the
closed-string condensation is intimately related to the emergence of the gauge
structure and the fermions. We have shown that the ground state of the exactly
soluble model can be constructed via the projective construction.  This
indicates that the projective construction is probably just a trick to
construct string condensed states.

\subsection{String operators and closed-string condensation}

\begin{figure}
\centerline{
\includegraphics[width=2.in]{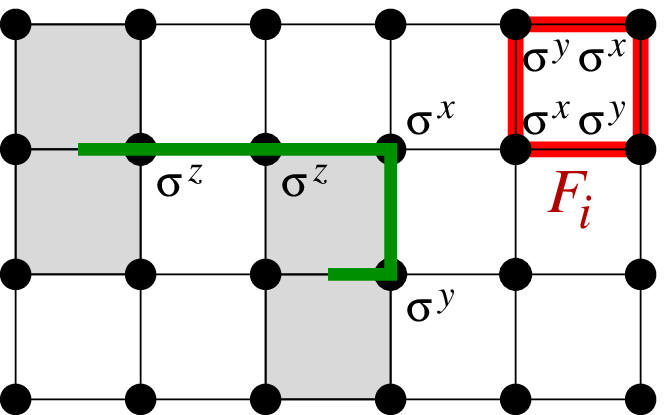}
}
\caption{
The string is formed by a curve connecting the midpoints of the neighboring
links. The string operator is form by the product of $\si^{x,y,z}_{\v i}$
for sites on the string. The operator $\hat F_{\v i}$ is also presented.
}
\label{t3string}
\end{figure}

First let us define a string $C$ as a curve that connect the midpoints
of neighboring links (see Fig. \ref{t3string}).
The string operator has the following form
\begin{equation}
\label{Wstr}
 W(C) = \prod_{n} \si^{l_n}_{\v i_n}
\end{equation}
where $\v i_n$ are sites on the string.  $l_n=z$ if the string does not turn
at site $\v i_n$.  $l_n=x$ or $y$ if the string makes a turn at site $\v i_n$.
$l_n=x$ if the turn forms a upper-right or lower-left corner.  $l_n=y$ if the
turn forms a lower-right or upper-left corner. (See Fig.  \ref{t3string}.)

By an explicit calculation, one can show that the closed string operator
defined above commute with the spin Hamiltonian \Eq{spinH}. So the ground
state $|\text{ground}\>$ of the spin Hamiltonian is also an eigenstate of the
closed-string operator. Since the eigenvalues of the closed string operator
are $\pm 1$. We have $ \<\text{ground}| W(C_{closed}) |\text{ground}\> =\pm
1$.  The ground state has a closed-string condensation.

We note that in a symmetry breaking state, the operator representing the order
parameter has a non-zero expectation value $\< \hat O_{\v i}\>=\phi$.  If we
define a string operator as the product of $\frac{\hat O_{\v i}}{\phi}$ along
a loop, the average of the string operator will be non-zero. However, we do
not regard the non-zero average of such a string operator to represent a
string condensation.  The real closed-string condensation must be
``unbreakable'', in the sense that we cannot break a  closed string operator
into several segments and find condensation in each segment.  The string
operator $\prod_{\v i} \frac{\hat O_{\v i}}{\phi}$ does not satisfy this
property. However, the string operator defined in \Eq{Wstr} is indeed
unbreakable.  This is because an open-string operator does not commute with
the spin Hamiltonian and does not condense $ \<\text{ground}| W(C_{open})
|\text{ground}\> =0 $. It is such a ``unbreakable'' closed-string
condensation indicates a new order in the ground state.

In terms of the Majorana fermions, the closed-string can be written as
\begin{equation*}
 W(C_{closed})=\prod_{\<\v i \v j\>} i\hat U_{\v i\v j}
\end{equation*}
where $\<\v i \v j\>$ are the nearest neighbor links that form the closed
string $C_{closed}$.
For a spin state obtained from the ansatz $s_{\v i\v j}$ (via the projection,
$\cP|\{s_{\v i\v j}\>$), we find that
\begin{equation*}
 \<\{s_{\v i\v j}\}|\cP W(C_{closed}) \cP|\{s_{\v i\v j}\}\>
= \prod_{\<\v i \v j\>} i s_{\v i\v j}
\end{equation*}
(note that for closed strings $[ W(C_{closed}), \cP]=0$).  Therefore the
closed string operator is nothing but the Wegner-Wilson loop operator
\cite{W7159,W7445} for the corresponding $Z_2$ gauge theory. We see an
intimate relation between the closed-string condensation and the emergence of
a gauge structure.

\subsection{Open string operators and $Z_2$ charges}

If the fluctuations of $s_{\v i\v j}$ represent $Z_2$ gauge fluctuations, what
are the $Z_2$ charges?  In a gauge theory, we know that a Wegner-Wilson
operator for an open string is not gauge invariant and is not a physical
operator (\ie is not an operator that acts within the physical Hilbert space).
The open-string operator defined in \Eq{Wstr} act with the  physical Hilbert
space and is a gauge invariant physical operator. Thus the Wegner-Wilson
operator for an open string does not correspond to our open string operator.
However, in the gauge theory, two charge operators connected by the
Wegner-Wilson operator, $\phi_{\v x_1}^\dag \phi_{\v x_2}e^{i\int_{\v x_1}^{\v
x_2}d\v x\cdot\v a}$, is gauge invariant. It is such a operator that
correspond to our open string operator.

In the gage theory, the Wegner-Wilson loop operator create a loop of electric
flux. The closed-string operator defined in \Eq{Wstr} has the same physical
meaning.  In the gage theory, the operator $\phi_{\v x_1}^\dag \phi_{\v
x_2}e^{i\int_{\v x_1}^{\v x_2}d\v x\cdot\v a}$ creates two charges connected
by a line of the electric flux.  Our open string operator defined in \Eq{Wstr}
does the same thing: it creates two $Z_2$ charges at its two ends and a
electric flux line connecting the two charges.

Due to the closed-string condensation, the string connecting the two $Z_2$
charges is unobservable and costs no energy.  Thus the open string does not
create an extended line-like object, it creates two point-like objects at its
ends. Those point-like objects are the $Z_2$ charges.  Despite the point-like
appearance, the $Z_2$ charges are intrinsically non-local. There is no way to
create a lone $Z_2$ charge.  Because of this, the $Z_2$ charge (and other
gauge charges) usually carry fractional quantum numbers.

\subsection{Statistics of the $Z_2$ charges}

What is the statistics of the $Z_2$ charges?  Usually, bosons are defined as
particles described by commuting operators and fermions as particles described
by anti-commuting operators. But this definition is too formal and is hard to
apply to our case.  The find a new way to calculate statistics, we need to
gain a more physical understanding of the difference between bosons and
fermions. 

Let us consider the following many-body hopping system.  The Hilbert space is
formed by a zero-particle state $|0\>$, one-particle states $|\v i_1\>$,
two-particle states $|\v i_1,\v i_2\>$, \etc, where $\v i_n$ labels the sites
in a lattice.  As an identical particle system, the state $|\v i_1,\v i_2,
...\>$ does not depend on the order of the indexes $\v i_1,\v i_2, ...$. For
example $|\v i_1,\v i_2\>=|\v i_2,\v i_1\>$.  There are no doubly-occupied
sites and we assume $|\v i_1,\v i_2, ...\>=0$ if $\v i_m=\v i_n$.

A hopping operator $\hat t_{\v i\v j}$ is defined as follows.  When $\hat
t_{\v i\v j}$ acts on state $|\v i_1,\v i_2, ...\>$, if there is a particle at
site $\v j$ but no particle at site $\v i$, then $\hat t_{\v i\v j}$ moves the
particle at site $\v j$ to site $\v i$ and multiplies a complex amplitude
$t(\v i,\v j;\v i_1,\v i_2,...)$ to the resulting state. Note that the
amplitude may depend on the locations of all the other particles.  The
Hamiltonian of our system is given by
\begin{equation*}
 H_{hop}=\sum_{\<\v i\v j\>} \hat t_{\v i\v j}
\end{equation*}
where the sum $\sum_{\<\v i\v j\>}$ is over a certain set of pairs $\<\v i\v
j\>$, such as nearest-neighbor pairs.  In order for the above Hamiltonian to
represent a local system we require that
\begin{equation*}
 [ \hat t_{\v i\v j}, \hat t_{\v k\v l}]=0
\end{equation*}
if $\v i$, $\v j$, $\v k$, and $\v l$ are all different. 

What does the hopping Hamiltonian $H_{hop}$ describe? A hard-core
boson system or a fermion system?  Whether a many-body hopping system is a
boson system or a fermion system (or even some other statistical systems) has
nothing to do with the Hilbert space.  The fact that the many-body states are
labeled by symmetric indexes (\eg $|\v i_1,\v i_2\>=|\v i_2,\v i_1\>$) does
not imply that the many-body system is a boson system. The statistics are
determined by the Hamiltonian $H_{hop}$.

Clearly, when the hopping amplitude $t(\v i,\v j;\v i_1,\v i_2,...)$ only
depends on $\v i$ and $\v j$, $t(\v i,\v j;\v i_1,\v i_2,...)=t(\v i,\v j)$,
the many-body hopping Hamiltonian will describe a hard-core boson system.  
The issue is under what condition the many-body hopping Hamiltonian describes
a fermion system.

This problem was solved in \Ref{LWsta}.  It was found that the many-body hopping
Hamiltonian describes a fermion system if the hopping operators satisfy
\index{fermion hopping algebra}
\begin{equation}
\label{fermhopalg}
 \hat t_{\v l\v k}
 \hat t_{\v i\v l}
 \hat t_{\v l\v j}
=-
 \hat t_{\v l\v j}
 \hat t_{\v i\v l}
 \hat t_{\v l\v k}
\end{equation}
for any three hopping operators 
$ \hat t_{\v l\v j}$,
$ \hat t_{\v i\v l}$, and
$ \hat t_{\v l\v k}$ with
$\v i$, $\v j$, $\v k$, and $\v l$ all being different.
(Note that the algebra has a structure $\hat t_1\hat t_2\hat t_3 
= -\hat t_3\hat t_2\hat t_1$.)

\begin{figure}[tb]
\centerline{
\includegraphics[width=2.0in]{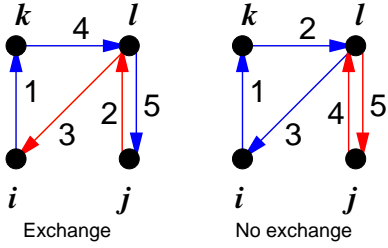}
}
\caption{ 
(a) The first way to arrange the five hops swaps the two particles.
(b) The second way to arrange the same five hops does not swap the 
two particles.
}
\label{Stat}
\end{figure}

To understand this result,
consider the state $|\v i,\v j,....\>$ with two particles at $\v i$, $\v j$, and
possibly other particles further away. We apply a set of five hopping operators
$\{\hat t_{\v j\v l}, \hat t_{\v l\v k}, \hat t_{\v i\v l}, \hat t_{\v l\v j},
\hat t_{\v k\v i} \}$ to the state
$|\v i,\v j,....\>$ but with different order (set Fig. \ref{Stat})
\begin{align*}
 \hat t_{\v j\v l}
 \hat t_{\v l\v k}
 \hat t_{\v i\v l}
 \hat t_{\v l\v j}
 \hat t_{\v k\v i} 
|\v i,\v j,....\> =& C_1 |\v i,\v j,....\> \nonumber\\
 \hat t_{\v j\v l}
 \hat t_{\v l\v j}
 \hat t_{\v i\v l}
 \hat t_{\v l\v k}
 \hat t_{\v k\v i} 
|\v i,\v j,....\> =& C_2 |\v i,\v j,....\> \nonumber\\
\end{align*}
where we have assumed that there are no particles at sites $\v k$ and $\v l$.
We note that after five hops we get back to the original state $|\v i,\v
j,....\>$ with additional phases $C_{1,2}$. However, from Fig. \ref{Stat}, we
see that the first way to arrange the five hops (Fig. \ref{Stat}a) swaps the
two particles at $\v i$ and $\v j$, while the second way (Fig.  \ref{Stat}b)
does not swap the two particles. Since the two hopping schemes use the same
set of five hops, the difference between $C_1$ and $C_2$ is due to exchanging
the two particles. Thus we require $C_1=-C_2$ in order for the many-body
hopping Hamiltonian to describe a fermion system.  Noting that the first and
the last hops are the same in the two hopping schemes, we find that $C_1=-C_2$
if the hopping operators satisfy \Eq{fermhopalg}.  \Eq{fermhopalg} serves as
an alternative definition of Fermi statistics if we do not want to use
anti-commuting algebra.

\begin{figure}[tb]
\centerline{
\includegraphics[width=2.0in]{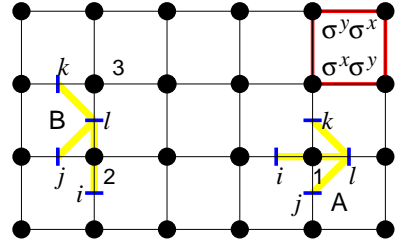}
}
\caption{ 
The $Z_2$ charge live on the links. The hopping of the $Z_2$ charges is
induced by the open string operator.
}
\label{fhopalg}
\end{figure}

For our spin model \Eq{spinH}, the open strings  end at the midpoint of the
links. So the $Z_2$ charges live on the links.  To apply the above result to
the $Z_2$ charges, we note that an open string operator connecting midpoints
$i$ and $j$ (see Fig. \ref{fhopalg}) play the role of a hopping operator
$\hat t_{ij}$.  Near the site 1 in Fig. \ref{fhopalg}, the hoping operators
between the midpoints $i$, $j$, $k$, and $l$ are given by
\begin{align*}
\hat t_{lj}=\si^y_1,\ \ \ \
\hat t_{il}=\si^z_1,\ \ \ \
\hat t_{lk}=\si^x_1.
\end{align*}
The fermion hopping algebra \Eq{fermhopalg} becomes $ \si^y_1 \si^z_1 \si^x_1
= - \si^x_1 \si^z_1 \si^y_1$ which is satisfied.  Near the site 2 and 3 in
Fig. \ref{fhopalg}, the hoping operators between the midpoints $i$, $j$, $k$,
and $l$ are given by
\begin{align*}
\hat t_{lj}=\si^y_2,\ \ \ \
\hat t_{il}=\si^z_2,\ \ \ \
\hat t_{lk}=\si^x_3
\end{align*}
The fermion hopping algebra becomes $\si^y_2 \si^z_2 \si^x_3 = - \si^x_3
\si^z_2 \si^y_2$ which is again satisfied.  We see that the hopping operators
of the $Z_2$ charges satisfy a fermion hopping algebra. So the $Z_2$ charges
are fermions.  Fermions can emerge in a pure bosonic model as ends of
condensed strings.

\section{Summary}

Symmetry breaking have dominated our understanding of phase and phase
transition for over 50 years. We now know that symmetry breaking cannot
describe all the possible orders that matter can have.  The study of the
quantum order \cite{Wqoslpub} and 
the associated condensation of string-nets \cite{Wqoem,LWstrnet} 
and other extended
objects suggest that a new world beyond symmetry breaking exists.  
The most striking picture from the new world is that gauge
bosons, fermions, and string-net condensation are just the different sides of
the same coin.  Or in other words, the string-net condensation provides a way
to unify gauge interactions and Fermi statistics! 
So far, we have only seen some small fragmented pieces of the
new world.  The exciting time is still ahead of us.  Comparing with our
understanding of symmetry breaking order, we need to understand the following
aspects of quantum order and the associated string-net condensation:\\
(1) We know that the group theory is the mathematical frame work behind the
symmetry breaking order. What is the mathematical frame work behind string
condensation? PSG only provides a partial answer to this question.
A recent work \cite{LWstrnet} 
suggests that tensor category theory may play the same role
in string-net condensed states as group theory in symmetry breaking states.
\\
(2) We know that crystal orders can be measured through X-ray diffraction.
How to measure different quantum orders associated with different string-net
condensations?\\
(3) We know that many material contain non-trivial symmetry breaking orders.
What material has string-net condensation and emergent gauge bosons and
fermions? (A believer can always say that we actually live inside one such
material. However, one needs more to convince a non-believer.)

I believe that the new world of quantum order is much richer than the world of
symmetry breaking order. Exploring the new world may represent a furture
direction in condensed matter research. It is hard to say what we will get
from the research in this direction. But one thing is sure: the more we
explore, the more are we fascinated by the endless richness of the nature.

This research is supported by NSF Grant No. DMR--01--23156
and by NSF-MRSEC Grant No. DMR--02--13282.

\bibliography{/home/wen/bib/wencross,/home/wen/bib/all,/home/wen/bib/publst}

\begin{thebibliography}{70}
\expandafter\ifx\csname natexlab\endcsname\relax\def\natexlab#1{#1}\fi
\expandafter\ifx\csname bibnamefont\endcsname\relax
  \def\bibnamefont#1{#1}\fi
\expandafter\ifx\csname bibfnamefont\endcsname\relax
  \def\bibfnamefont#1{#1}\fi
\expandafter\ifx\csname citenamefont\endcsname\relax
  \def\citenamefont#1{#1}\fi
\expandafter\ifx\csname url\endcsname\relax
  \def\url#1{\texttt{#1}}\fi
\expandafter\ifx\csname urlprefix\endcsname\relax\def\urlprefix{URL }\fi
\providecommand{\bibinfo}[2]{#2}
\providecommand{\eprint}[2][]{\url{#2}}

\bibitem[{\citenamefont{Nambu}(1960)}]{N6080}
\bibinfo{author}{\bibfnamefont{Y.}~\bibnamefont{Nambu}},
  \bibinfo{journal}{Phys. Rev. Lett.} \textbf{\bibinfo{volume}{4}},
  \bibinfo{pages}{380} (\bibinfo{year}{1960}).

\bibitem[{\citenamefont{Goldstone}(1961)}]{G6154}
\bibinfo{author}{\bibfnamefont{J.}~\bibnamefont{Goldstone}},
  \bibinfo{journal}{Nuovo Cimento} \textbf{\bibinfo{volume}{19}},
  \bibinfo{pages}{154} (\bibinfo{year}{1961}).

\bibitem[{\citenamefont{Landau and Lifschitz}(1958)}]{LanL58}
\bibinfo{author}{\bibfnamefont{L.~D.} \bibnamefont{Landau}} \bibnamefont{and}
  \bibinfo{author}{\bibfnamefont{E.~M.} \bibnamefont{Lifschitz}},
  \emph{\bibinfo{title}{Statistical Physics - Course of Theoretical Physics Vol
  5}} (\bibinfo{publisher}{Pergamon}, \bibinfo{address}{London},
  \bibinfo{year}{1958}).

\bibitem[{\citenamefont{Landau}(1937)}]{L3726}
\bibinfo{author}{\bibfnamefont{L.~D.} \bibnamefont{Landau}},
  \bibinfo{journal}{Phys. Z. Sowjetunion} \textbf{\bibinfo{volume}{11}},
  \bibinfo{pages}{26} (\bibinfo{year}{1937}).

\bibitem[{\citenamefont{Tsui et~al.}(1982)\citenamefont{Tsui, Stormer, and
  Gossard}}]{TSG8259}
\bibinfo{author}{\bibfnamefont{D.~C.} \bibnamefont{Tsui}},
  \bibinfo{author}{\bibfnamefont{H.~L.} \bibnamefont{Stormer}},
  \bibnamefont{and} \bibinfo{author}{\bibfnamefont{A.~C.}
  \bibnamefont{Gossard}}, \bibinfo{journal}{Phys. Rev. Lett.}
  \textbf{\bibinfo{volume}{48}}, \bibinfo{pages}{1559} (\bibinfo{year}{1982}).

\bibitem[{\citenamefont{Laughlin}(1983)}]{L8395}
\bibinfo{author}{\bibfnamefont{R.~B.} \bibnamefont{Laughlin}},
  \bibinfo{journal}{Phys. Rev. Lett.} \textbf{\bibinfo{volume}{50}},
  \bibinfo{pages}{1395} (\bibinfo{year}{1983}).

\bibitem[{\citenamefont{Wen}(1995)}]{Wtoprev}
\bibinfo{author}{\bibfnamefont{X.-G.} \bibnamefont{Wen}},
  \bibinfo{journal}{Advances in Physics} \textbf{\bibinfo{volume}{44}},
  \bibinfo{pages}{405} (\bibinfo{year}{1995}).

\bibitem[{\citenamefont{Haldane and Rezayi}(1985)}]{HR8529}
\bibinfo{author}{\bibfnamefont{F.~D.~M.} \bibnamefont{Haldane}}
  \bibnamefont{and} \bibinfo{author}{\bibfnamefont{E.~H.}
  \bibnamefont{Rezayi}}, \bibinfo{journal}{Phys. Rev. B}
  \textbf{\bibinfo{volume}{31}}, \bibinfo{pages}{2529} (\bibinfo{year}{1985}).

\bibitem[{\citenamefont{Wen and Niu}(1990)}]{WNtop}
\bibinfo{author}{\bibfnamefont{X.-G.} \bibnamefont{Wen}} \bibnamefont{and}
  \bibinfo{author}{\bibfnamefont{Q.}~\bibnamefont{Niu}},
  \bibinfo{journal}{Phys. Rev. B} \textbf{\bibinfo{volume}{41}},
  \bibinfo{pages}{9377} (\bibinfo{year}{1990}).

\bibitem[{\citenamefont{Wen}(1990)}]{Wrig}
\bibinfo{author}{\bibfnamefont{X.-G.} \bibnamefont{Wen}},
  \bibinfo{journal}{Int. J. Mod. Phys. B} \textbf{\bibinfo{volume}{4}},
  \bibinfo{pages}{239} (\bibinfo{year}{1990}).

\bibitem[{\citenamefont{Arovas et~al.}(1984)\citenamefont{Arovas, Schrieffer,
  and Wilczek}}]{ASW8422}
\bibinfo{author}{\bibfnamefont{D.}~\bibnamefont{Arovas}},
  \bibinfo{author}{\bibfnamefont{J.~R.} \bibnamefont{Schrieffer}},
  \bibnamefont{and} \bibinfo{author}{\bibfnamefont{F.}~\bibnamefont{Wilczek}},
  \bibinfo{journal}{Phys. Rev. Lett.} \textbf{\bibinfo{volume}{53}},
  \bibinfo{pages}{722} (\bibinfo{year}{1984}).

\bibitem[{\citenamefont{Halperin}(1982)}]{H8285}
\bibinfo{author}{\bibfnamefont{B.~I.} \bibnamefont{Halperin}},
  \bibinfo{journal}{Phys. Rev. B} \textbf{\bibinfo{volume}{25}},
  \bibinfo{pages}{2185} (\bibinfo{year}{1982}).

\bibitem[{\citenamefont{Wen}(1992)}]{Wedgerev}
\bibinfo{author}{\bibfnamefont{X.-G.} \bibnamefont{Wen}},
  \bibinfo{journal}{Int. J. Mod. Phys. B} \textbf{\bibinfo{volume}{6}},
  \bibinfo{pages}{1711} (\bibinfo{year}{1992}).

\bibitem[{\citenamefont{Witten}(1989)}]{W8951}
\bibinfo{author}{\bibfnamefont{E.}~\bibnamefont{Witten}},
  \bibinfo{journal}{Comm. Math. Phys.} \textbf{\bibinfo{volume}{121}},
  \bibinfo{pages}{351} (\bibinfo{year}{1989}).

\bibitem[{\citenamefont{Kitaev}(2003)}]{K032}
\bibinfo{author}{\bibfnamefont{A.~Y.} \bibnamefont{Kitaev}},
  \bibinfo{journal}{Ann. Phys. (N.Y.)} \textbf{\bibinfo{volume}{303}},
  \bibinfo{pages}{2} (\bibinfo{year}{2003}).

\bibitem[{\citenamefont{Wen}(2002{\natexlab{a}})}]{Wqoslpub}
\bibinfo{author}{\bibfnamefont{X.-G.} \bibnamefont{Wen}},
  \bibinfo{journal}{Phys. Rev. B} \textbf{\bibinfo{volume}{65}},
  \bibinfo{pages}{165113} (\bibinfo{year}{2002}{\natexlab{a}}).

\bibitem[{\citenamefont{Lifshitz}(1960)}]{L6030}
\bibinfo{author}{\bibfnamefont{I.~M.} \bibnamefont{Lifshitz}},
  \bibinfo{journal}{Sov. Phys. JETP} \textbf{\bibinfo{volume}{11}},
  \bibinfo{pages}{1130} (\bibinfo{year}{1960}).

\bibitem[{\citenamefont{Wen}(2003{\natexlab{a}})}]{Walight}
\bibinfo{author}{\bibfnamefont{X.-G.} \bibnamefont{Wen}},
  \bibinfo{journal}{Phys. Rev. B} \textbf{\bibinfo{volume}{68}},
  \bibinfo{pages}{115413} (\bibinfo{year}{2003}{\natexlab{a}}).

\bibitem[{\citenamefont{Levin and Wen}(2003)}]{LWsta}
\bibinfo{author}{\bibfnamefont{M.}~\bibnamefont{Levin}} \bibnamefont{and}
  \bibinfo{author}{\bibfnamefont{X.-G.} \bibnamefont{Wen}},
  \bibinfo{journal}{Phys. Rev. B} \textbf{\bibinfo{volume}{67}},
  \bibinfo{pages}{245316} (\bibinfo{year}{2003}).

\bibitem[{\citenamefont{Freedman et~al.}(2003)\citenamefont{Freedman, Nayak,
  Shtengel, Walker, and Wang}}]{FNS0311}
\bibinfo{author}{\bibfnamefont{M.}~\bibnamefont{Freedman}},
  \bibinfo{author}{\bibfnamefont{C.}~\bibnamefont{Nayak}},
  \bibinfo{author}{\bibfnamefont{K.}~\bibnamefont{Shtengel}},
  \bibinfo{author}{\bibfnamefont{K.}~\bibnamefont{Walker}}, \bibnamefont{and}
  \bibinfo{author}{\bibfnamefont{Z.}~\bibnamefont{Wang}},
  \bibinfo{journal}{cond-mat} p. \bibinfo{pages}{0307511}
  (\bibinfo{year}{2003}).

\bibitem[{\citenamefont{Banks et~al.}(1977)\citenamefont{Banks, Myerson, and
  Kogut}}]{BMK7793}
\bibinfo{author}{\bibfnamefont{T.}~\bibnamefont{Banks}},
  \bibinfo{author}{\bibfnamefont{R.}~\bibnamefont{Myerson}}, \bibnamefont{and}
  \bibinfo{author}{\bibfnamefont{J.~B.} \bibnamefont{Kogut}},
  \bibinfo{journal}{Nucl. Phys. B} \textbf{\bibinfo{volume}{129}},
  \bibinfo{pages}{493} (\bibinfo{year}{1977}).

\bibitem[{\citenamefont{Wegner}(1971)}]{W7159}
\bibinfo{author}{\bibfnamefont{F.}~\bibnamefont{Wegner}}, \bibinfo{journal}{J.
  Math. Phys.} \textbf{\bibinfo{volume}{12}}, \bibinfo{pages}{2259}
  (\bibinfo{year}{1971}).

\bibitem[{\citenamefont{Kogut and Susskind}(1975)}]{KS7595}
\bibinfo{author}{\bibfnamefont{J.}~\bibnamefont{Kogut}} \bibnamefont{and}
  \bibinfo{author}{\bibfnamefont{L.}~\bibnamefont{Susskind}},
  \bibinfo{journal}{Phys. Rev. D} \textbf{\bibinfo{volume}{11}},
  \bibinfo{pages}{395} (\bibinfo{year}{1975}).

\bibitem[{\citenamefont{Levin and Wen}(2004)}]{LWstrnet}
\bibinfo{author}{\bibfnamefont{M.}~\bibnamefont{Levin}} \bibnamefont{and}
  \bibinfo{author}{\bibfnamefont{X.-G.} \bibnamefont{Wen}},
  \bibinfo{journal}{cond-mat/0404617}  (\bibinfo{year}{2004}).

\bibitem[{\citenamefont{Wen and Wu}(1993)}]{WWtran}
\bibinfo{author}{\bibfnamefont{X.-G.} \bibnamefont{Wen}} \bibnamefont{and}
  \bibinfo{author}{\bibfnamefont{Y.-S.} \bibnamefont{Wu}},
  \bibinfo{journal}{Phys. Rev. Lett.} \textbf{\bibinfo{volume}{70}},
  \bibinfo{pages}{1501} (\bibinfo{year}{1993}).

\bibitem[{\citenamefont{Chen et~al.}(1993)\citenamefont{Chen, Fisher, and
  Wu}}]{CFW9349}
\bibinfo{author}{\bibfnamefont{W.}~\bibnamefont{Chen}},
  \bibinfo{author}{\bibfnamefont{M.~P.~A.} \bibnamefont{Fisher}},
  \bibnamefont{and} \bibinfo{author}{\bibfnamefont{Y.-S.} \bibnamefont{Wu}},
  \bibinfo{journal}{Phys. Rev. B} \textbf{\bibinfo{volume}{48}},
  \bibinfo{pages}{13749} (\bibinfo{year}{1993}).

\bibitem[{\citenamefont{Senthil et~al.}(1999)\citenamefont{Senthil, Marston,
  and Fisher}}]{SMF9945}
\bibinfo{author}{\bibfnamefont{T.}~\bibnamefont{Senthil}},
  \bibinfo{author}{\bibfnamefont{J.~B.} \bibnamefont{Marston}},
  \bibnamefont{and} \bibinfo{author}{\bibfnamefont{M.~P.~A.}
  \bibnamefont{Fisher}}, \bibinfo{journal}{Phys. Rev. B}
  \textbf{\bibinfo{volume}{60}}, \bibinfo{pages}{4245} (\bibinfo{year}{1999}).

\bibitem[{\citenamefont{Read and Green}(2000)}]{RG0067}
\bibinfo{author}{\bibfnamefont{N.}~\bibnamefont{Read}} \bibnamefont{and}
  \bibinfo{author}{\bibfnamefont{D.}~\bibnamefont{Green}},
  \bibinfo{journal}{Phys. Rev. B} \textbf{\bibinfo{volume}{61}},
  \bibinfo{pages}{10267} (\bibinfo{year}{2000}).

\bibitem[{\citenamefont{Wen}(2000)}]{Wctpt}
\bibinfo{author}{\bibfnamefont{X.-G.} \bibnamefont{Wen}},
  \bibinfo{journal}{Phys. Rev. Lett.} \textbf{\bibinfo{volume}{84}},
  \bibinfo{pages}{3950} (\bibinfo{year}{2000}).

\bibitem[{\citenamefont{Wen}(2002{\natexlab{b}})}]{Wlight}
\bibinfo{author}{\bibfnamefont{X.-G.} \bibnamefont{Wen}},
  \bibinfo{journal}{Phys. Rev. Lett.} \textbf{\bibinfo{volume}{88}},
  \bibinfo{pages}{11602} (\bibinfo{year}{2002}{\natexlab{b}}).

\bibitem[{\citenamefont{Wen and Zee}(2002)}]{WZqoind}
\bibinfo{author}{\bibfnamefont{X.-G.} \bibnamefont{Wen}} \bibnamefont{and}
  \bibinfo{author}{\bibfnamefont{A.}~\bibnamefont{Zee}},
  \bibinfo{journal}{Phys. Rev. B} \textbf{\bibinfo{volume}{66}},
  \bibinfo{pages}{235110} (\bibinfo{year}{2002}).

\bibitem[{\citenamefont{Moessner and Sondhi}(2001)}]{MS0181}
\bibinfo{author}{\bibfnamefont{R.}~\bibnamefont{Moessner}} \bibnamefont{and}
  \bibinfo{author}{\bibfnamefont{S.~L.} \bibnamefont{Sondhi}},
  \bibinfo{journal}{Phys. Rev. Lett.} \textbf{\bibinfo{volume}{86}},
  \bibinfo{pages}{1881} (\bibinfo{year}{2001}).

\bibitem[{\citenamefont{Read and Sachdev}(1991)}]{RS9173}
\bibinfo{author}{\bibfnamefont{N.}~\bibnamefont{Read}} \bibnamefont{and}
  \bibinfo{author}{\bibfnamefont{S.}~\bibnamefont{Sachdev}},
  \bibinfo{journal}{Phys. Rev. Lett.} \textbf{\bibinfo{volume}{66}},
  \bibinfo{pages}{1773} (\bibinfo{year}{1991}).

\bibitem[{\citenamefont{Wen}(1991{\natexlab{a}})}]{Wsrvb}
\bibinfo{author}{\bibfnamefont{X.-G.} \bibnamefont{Wen}},
  \bibinfo{journal}{Phys. Rev. B} \textbf{\bibinfo{volume}{44}},
  \bibinfo{pages}{2664} (\bibinfo{year}{1991}{\natexlab{a}}).

\bibitem[{\citenamefont{Kalmeyer and Laughlin}(1987)}]{KL8795}
\bibinfo{author}{\bibfnamefont{V.}~\bibnamefont{Kalmeyer}} \bibnamefont{and}
  \bibinfo{author}{\bibfnamefont{R.~B.} \bibnamefont{Laughlin}},
  \bibinfo{journal}{Phys. Rev. Lett.} \textbf{\bibinfo{volume}{59}},
  \bibinfo{pages}{2095} (\bibinfo{year}{1987}).

\bibitem[{\citenamefont{Wen et~al.}(1989)\citenamefont{Wen, Wilczek, and
  Zee}}]{WWZcsp}
\bibinfo{author}{\bibfnamefont{X.-G.} \bibnamefont{Wen}},
  \bibinfo{author}{\bibfnamefont{F.}~\bibnamefont{Wilczek}}, \bibnamefont{and}
  \bibinfo{author}{\bibfnamefont{A.}~\bibnamefont{Zee}},
  \bibinfo{journal}{Phys. Rev. B} \textbf{\bibinfo{volume}{39}},
  \bibinfo{pages}{11413} (\bibinfo{year}{1989}).

\bibitem[{\citenamefont{Foerster et~al.}(1980)\citenamefont{Foerster, Nielsen,
  and Ninomiya}}]{FNN8035}
\bibinfo{author}{\bibfnamefont{D.}~\bibnamefont{Foerster}},
  \bibinfo{author}{\bibfnamefont{H.~B.} \bibnamefont{Nielsen}},
  \bibnamefont{and} \bibinfo{author}{\bibfnamefont{M.}~\bibnamefont{Ninomiya}},
  \bibinfo{journal}{Phys. Lett. B} \textbf{\bibinfo{volume}{94}},
  \bibinfo{pages}{135} (\bibinfo{year}{1980}).

\bibitem[{\citenamefont{Motrunich and Senthil}(2002)}]{MS0204}
\bibinfo{author}{\bibfnamefont{O.~I.} \bibnamefont{Motrunich}}
  \bibnamefont{and} \bibinfo{author}{\bibfnamefont{T.}~\bibnamefont{Senthil}},
  \bibinfo{journal}{Phys. Rev. Lett.} \textbf{\bibinfo{volume}{89}},
  \bibinfo{pages}{277004} (\bibinfo{year}{2002}).

\bibitem[{\citenamefont{Moessner and Sondhi}(2003)}]{MS0312}
\bibinfo{author}{\bibfnamefont{R.}~\bibnamefont{Moessner}} \bibnamefont{and}
  \bibinfo{author}{\bibfnamefont{S.~L.} \bibnamefont{Sondhi}},
  \bibinfo{journal}{Phys. Rev. B} \textbf{\bibinfo{volume}{68}},
  \bibinfo{pages}{184512} (\bibinfo{year}{2003}).

\bibitem[{\citenamefont{Hermele
  et~al.}(2004{\natexlab{a}})\citenamefont{Hermele, Fisher, and
  Balents}}]{HFB0404}
\bibinfo{author}{\bibfnamefont{M.}~\bibnamefont{Hermele}},
  \bibinfo{author}{\bibfnamefont{M.~P.~A.} \bibnamefont{Fisher}},
  \bibnamefont{and} \bibinfo{author}{\bibfnamefont{L.}~\bibnamefont{Balents}},
  \bibinfo{journal}{Phys. Rev. B} \textbf{\bibinfo{volume}{69}},
  \bibinfo{pages}{064404} (\bibinfo{year}{2004}{\natexlab{a}}).

\bibitem[{\citenamefont{Wen}(2003{\natexlab{b}})}]{Wqoem}
\bibinfo{author}{\bibfnamefont{X.-G.} \bibnamefont{Wen}},
  \bibinfo{journal}{Phys. Rev. D} \textbf{\bibinfo{volume}{68}},
  \bibinfo{pages}{065003} (\bibinfo{year}{2003}{\natexlab{b}}).

\bibitem[{\citenamefont{Blok and Wen}(1990)}]{BWkmat2}
\bibinfo{author}{\bibfnamefont{B.}~\bibnamefont{Blok}} \bibnamefont{and}
  \bibinfo{author}{\bibfnamefont{X.-G.} \bibnamefont{Wen}},
  \bibinfo{journal}{Phys. Rev. B} \textbf{\bibinfo{volume}{42}},
  \bibinfo{pages}{8145} (\bibinfo{year}{1990}).

\bibitem[{\citenamefont{Read}(1990)}]{R9002}
\bibinfo{author}{\bibfnamefont{N.}~\bibnamefont{Read}}, \bibinfo{journal}{Phys.
  Rev. Lett.} \textbf{\bibinfo{volume}{65}}, \bibinfo{pages}{1502}
  (\bibinfo{year}{1990}).

\bibitem[{\citenamefont{Fr{\"o}hlich and Kerler}(1991)}]{FK9169}
\bibinfo{author}{\bibfnamefont{J.}~\bibnamefont{Fr{\"o}hlich}}
  \bibnamefont{and} \bibinfo{author}{\bibfnamefont{T.}~\bibnamefont{Kerler}},
  \bibinfo{journal}{Nucl. Phys. B} \textbf{\bibinfo{volume}{354}},
  \bibinfo{pages}{369} (\bibinfo{year}{1991}).

\bibitem[{\citenamefont{Rokhsar and Kivelson}(1988)}]{RK8876}
\bibinfo{author}{\bibfnamefont{D.~S.} \bibnamefont{Rokhsar}} \bibnamefont{and}
  \bibinfo{author}{\bibfnamefont{S.~A.} \bibnamefont{Kivelson}},
  \bibinfo{journal}{Phys. Rev. Lett.} \textbf{\bibinfo{volume}{61}},
  \bibinfo{pages}{2376} (\bibinfo{year}{1988}).

\bibitem[{\citenamefont{Read and Chakraborty}(1989)}]{RC8933}
\bibinfo{author}{\bibfnamefont{N.}~\bibnamefont{Read}} \bibnamefont{and}
  \bibinfo{author}{\bibfnamefont{B.}~\bibnamefont{Chakraborty}},
  \bibinfo{journal}{Phys. Rev. B} \textbf{\bibinfo{volume}{40}},
  \bibinfo{pages}{7133} (\bibinfo{year}{1989}).

\bibitem[{\citenamefont{Ardonne et~al.}(2004)\citenamefont{Ardonne, Fendley,
  and Fradkin}}]{AFF0493}
\bibinfo{author}{\bibfnamefont{E.}~\bibnamefont{Ardonne}},
  \bibinfo{author}{\bibfnamefont{P.}~\bibnamefont{Fendley}}, \bibnamefont{and}
  \bibinfo{author}{\bibfnamefont{E.}~\bibnamefont{Fradkin}},
  \bibinfo{journal}{Annals Phys.} \textbf{\bibinfo{volume}{310}},
  \bibinfo{pages}{493} (\bibinfo{year}{2004}).

\bibitem[{\citenamefont{Senthil and Fisher}(2000)}]{SF0050}
\bibinfo{author}{\bibfnamefont{T.}~\bibnamefont{Senthil}} \bibnamefont{and}
  \bibinfo{author}{\bibfnamefont{M.~P.~A.} \bibnamefont{Fisher}},
  \bibinfo{journal}{Phys. Rev. B} \textbf{\bibinfo{volume}{62}},
  \bibinfo{pages}{7850} (\bibinfo{year}{2000}).

\bibitem[{\citenamefont{Sachdev and Park}(2002)}]{SP0258}
\bibinfo{author}{\bibfnamefont{S.}~\bibnamefont{Sachdev}} \bibnamefont{and}
  \bibinfo{author}{\bibfnamefont{K.}~\bibnamefont{Park}},
  \bibinfo{journal}{Annals of Physics (N.Y.)} \textbf{\bibinfo{volume}{298}},
  \bibinfo{pages}{58} (\bibinfo{year}{2002}).

\bibitem[{\citenamefont{Balents et~al.}(2002)\citenamefont{Balents, Fisher, and
  Girvin}}]{BFG0212}
\bibinfo{author}{\bibfnamefont{L.}~\bibnamefont{Balents}},
  \bibinfo{author}{\bibfnamefont{M.~P.~A.} \bibnamefont{Fisher}},
  \bibnamefont{and} \bibinfo{author}{\bibfnamefont{S.~M.}
  \bibnamefont{Girvin}}, \bibinfo{journal}{Phys. Rev. B}
  \textbf{\bibinfo{volume}{65}}, \bibinfo{pages}{224412}
  (\bibinfo{year}{2002}).

\bibitem[{\citenamefont{Wen}(1991{\natexlab{b}})}]{Wtopcs}
\bibinfo{author}{\bibfnamefont{X.-G.} \bibnamefont{Wen}},
  \bibinfo{journal}{Int. J. Mod. Phys. B} \textbf{\bibinfo{volume}{5}},
  \bibinfo{pages}{1641} (\bibinfo{year}{1991}{\natexlab{b}}).

\bibitem[{\citenamefont{Hansson et~al.}(2004)\citenamefont{Hansson, Oganesyan,
  and Sondhi}}]{HOS0427}
\bibinfo{author}{\bibfnamefont{T.~H.} \bibnamefont{Hansson}},
  \bibinfo{author}{\bibfnamefont{V.}~\bibnamefont{Oganesyan}},
  \bibnamefont{and} \bibinfo{author}{\bibfnamefont{S.~L.}
  \bibnamefont{Sondhi}}, \bibinfo{journal}{cond-mat/0404327}
  (\bibinfo{year}{2004}).

\bibitem[{\citenamefont{Kogut}(1979)}]{K7959}
\bibinfo{author}{\bibfnamefont{J.~B.} \bibnamefont{Kogut}},
  \bibinfo{journal}{Rev. Mod. Phys.} \textbf{\bibinfo{volume}{51}},
  \bibinfo{pages}{659} (\bibinfo{year}{1979}).

\bibitem[{\citenamefont{Ioffe et~al.}(2002)\citenamefont{Ioffe, Feigel'man,
  Ioselevich, Ivanov, Troyer, and Blatter}}]{IFI0203}
\bibinfo{author}{\bibfnamefont{L.~B.} \bibnamefont{Ioffe}},
  \bibinfo{author}{\bibfnamefont{M.~V.} \bibnamefont{Feigel'man}},
  \bibinfo{author}{\bibfnamefont{A.}~\bibnamefont{Ioselevich}},
  \bibinfo{author}{\bibfnamefont{D.}~\bibnamefont{Ivanov}},
  \bibinfo{author}{\bibfnamefont{M.}~\bibnamefont{Troyer}}, \bibnamefont{and}
  \bibinfo{author}{\bibfnamefont{G.}~\bibnamefont{Blatter}},
  \bibinfo{journal}{Nature} \textbf{\bibinfo{volume}{415}},
  \bibinfo{pages}{503} (\bibinfo{year}{2002}).

\bibitem[{\citenamefont{Freedman et~al.}(2002)\citenamefont{Freedman, Larsen,
  and Wang}}]{FLZ0205}
\bibinfo{author}{\bibfnamefont{M.}~\bibnamefont{Freedman}},
  \bibinfo{author}{\bibfnamefont{M.}~\bibnamefont{Larsen}}, \bibnamefont{and}
  \bibinfo{author}{\bibfnamefont{Z.}~\bibnamefont{Wang}},
  \bibinfo{journal}{Commun. Math. Phys.} \textbf{\bibinfo{volume}{227}},
  \bibinfo{pages}{605} (\bibinfo{year}{2002}).

\bibitem[{\citenamefont{Ginzburg and Landau}(1950)}]{GL5064}
\bibinfo{author}{\bibfnamefont{V.~L.} \bibnamefont{Ginzburg}} \bibnamefont{and}
  \bibinfo{author}{\bibfnamefont{L.~D.} \bibnamefont{Landau}},
  \bibinfo{journal}{Zh. Ekaper. Teoret. Fiz.} \textbf{\bibinfo{volume}{20}},
  \bibinfo{pages}{1064} (\bibinfo{year}{1950}).

\bibitem[{\citenamefont{Affleck and Marston}(1988)}]{AM8874}
\bibinfo{author}{\bibfnamefont{I.}~\bibnamefont{Affleck}} \bibnamefont{and}
  \bibinfo{author}{\bibfnamefont{J.~B.} \bibnamefont{Marston}},
  \bibinfo{journal}{Phys. Rev. B} \textbf{\bibinfo{volume}{37}},
  \bibinfo{pages}{3774} (\bibinfo{year}{1988}).

\bibitem[{\citenamefont{Baskaran et~al.}(1987)\citenamefont{Baskaran, Zou, and
  Anderson}}]{BZA8773}
\bibinfo{author}{\bibfnamefont{G.}~\bibnamefont{Baskaran}},
  \bibinfo{author}{\bibfnamefont{Z.}~\bibnamefont{Zou}}, \bibnamefont{and}
  \bibinfo{author}{\bibfnamefont{P.~W.} \bibnamefont{Anderson}},
  \bibinfo{journal}{Solid State Comm.} \textbf{\bibinfo{volume}{63}},
  \bibinfo{pages}{973} (\bibinfo{year}{1987}).

\bibitem[{\citenamefont{Kotliar and Liu}(1988)}]{KL8842}
\bibinfo{author}{\bibfnamefont{G.}~\bibnamefont{Kotliar}} \bibnamefont{and}
  \bibinfo{author}{\bibfnamefont{J.}~\bibnamefont{Liu}},
  \bibinfo{journal}{Phys. Rev. B} \textbf{\bibinfo{volume}{38}},
  \bibinfo{pages}{5142} (\bibinfo{year}{1988}).

\bibitem[{\citenamefont{Marshall et~al.}(1996)\citenamefont{Marshall, Dessau,
  Loeser, Park, Matsuura, Eckstein, Bozovic, Fournier, Kapitulnik, Spicer
  et~al.}}]{MDL9641}
\bibinfo{author}{\bibfnamefont{D.~S.} \bibnamefont{Marshall}},
  \bibinfo{author}{\bibfnamefont{D.~S.} \bibnamefont{Dessau}},
  \bibinfo{author}{\bibfnamefont{A.~G.} \bibnamefont{Loeser}},
  \bibinfo{author}{\bibfnamefont{C.-H.} \bibnamefont{Park}},
  \bibinfo{author}{\bibfnamefont{A.~Y.} \bibnamefont{Matsuura}},
  \bibinfo{author}{\bibfnamefont{J.~N.} \bibnamefont{Eckstein}},
  \bibinfo{author}{\bibfnamefont{I.}~\bibnamefont{Bozovic}},
  \bibinfo{author}{\bibfnamefont{P.}~\bibnamefont{Fournier}},
  \bibinfo{author}{\bibfnamefont{A.}~\bibnamefont{Kapitulnik}},
  \bibinfo{author}{\bibfnamefont{W.~E.} \bibnamefont{Spicer}},
  \bibnamefont{et~al.}, \bibinfo{journal}{Phys. Rev. Lett.}
  \textbf{\bibinfo{volume}{76}}, \bibinfo{pages}{4841} (\bibinfo{year}{1996}).

\bibitem[{\citenamefont{Anderson}(1963)}]{A6339}
\bibinfo{author}{\bibfnamefont{P.~W.} \bibnamefont{Anderson}},
  \bibinfo{journal}{Phys. Rev.} \textbf{\bibinfo{volume}{130}},
  \bibinfo{pages}{439} (\bibinfo{year}{1963}).

\bibitem[{\citenamefont{Higgs}(1964)}]{H6432}
\bibinfo{author}{\bibfnamefont{P.~W.} \bibnamefont{Higgs}},
  \bibinfo{journal}{Phys. Rev. Lett.} \textbf{\bibinfo{volume}{12}},
  \bibinfo{pages}{132} (\bibinfo{year}{1964}).

\bibitem[{\citenamefont{Rantner and Wen}(2002)}]{RWspin}
\bibinfo{author}{\bibfnamefont{W.}~\bibnamefont{Rantner}} \bibnamefont{and}
  \bibinfo{author}{\bibfnamefont{X.-G.} \bibnamefont{Wen}},
  \bibinfo{journal}{Phys. Rev. B} \textbf{\bibinfo{volume}{66}},
  \bibinfo{pages}{144501} (\bibinfo{year}{2002}).

\bibitem[{\citenamefont{Hermele
  et~al.}(2004{\natexlab{b}})\citenamefont{Hermele, Senthil, Fisher, Lee,
  Nagaosa, and Wen}}]{HSF0451}
\bibinfo{author}{\bibfnamefont{M.}~\bibnamefont{Hermele}},
  \bibinfo{author}{\bibfnamefont{T.}~\bibnamefont{Senthil}},
  \bibinfo{author}{\bibfnamefont{M.~P.~A.} \bibnamefont{Fisher}},
  \bibinfo{author}{\bibfnamefont{P.~A.} \bibnamefont{Lee}},
  \bibinfo{author}{\bibfnamefont{N.}~\bibnamefont{Nagaosa}}, \bibnamefont{and}
  \bibinfo{author}{\bibfnamefont{X.-G.} \bibnamefont{Wen}},
  \bibinfo{journal}{cond-mat/0404751}  (\bibinfo{year}{2004}{\natexlab{b}}).

\bibitem[{\citenamefont{Ioffe and Larkin}(1989)}]{IL8988}
\bibinfo{author}{\bibfnamefont{L.}~\bibnamefont{Ioffe}} \bibnamefont{and}
  \bibinfo{author}{\bibfnamefont{A.}~\bibnamefont{Larkin}},
  \bibinfo{journal}{Phys. Rev. B} \textbf{\bibinfo{volume}{39}},
  \bibinfo{pages}{8988} (\bibinfo{year}{1989}).

\bibitem[{\citenamefont{Kivelson et~al.}(1987)\citenamefont{Kivelson, Rokhsar,
  and Sethna}}]{KRS8765}
\bibinfo{author}{\bibfnamefont{S.~A.} \bibnamefont{Kivelson}},
  \bibinfo{author}{\bibfnamefont{D.~S.} \bibnamefont{Rokhsar}},
  \bibnamefont{and} \bibinfo{author}{\bibfnamefont{J.~P.}
  \bibnamefont{Sethna}}, \bibinfo{journal}{Phys. Rev. B}
  \textbf{\bibinfo{volume}{35}}, \bibinfo{pages}{8865} (\bibinfo{year}{1987}).

\bibitem[{\citenamefont{Wen}(1999)}]{Wpcon}
\bibinfo{author}{\bibfnamefont{X.-G.} \bibnamefont{Wen}},
  \bibinfo{journal}{Phys. Rev. B} \textbf{\bibinfo{volume}{60}},
  \bibinfo{pages}{8827} (\bibinfo{year}{1999}).

\bibitem[{\citenamefont{Jain}(1991)}]{J9153}
\bibinfo{author}{\bibfnamefont{J.~K.} \bibnamefont{Jain}},
  \bibinfo{journal}{Phys. Rev. B} \textbf{\bibinfo{volume}{41}},
  \bibinfo{pages}{7653} (\bibinfo{year}{1991}).

\bibitem[{\citenamefont{Wen}(2003{\natexlab{c}})}]{Wqoexct}
\bibinfo{author}{\bibfnamefont{X.-G.} \bibnamefont{Wen}},
  \bibinfo{journal}{Phys. Rev. Lett.} \textbf{\bibinfo{volume}{90}},
  \bibinfo{pages}{016803} (\bibinfo{year}{2003}{\natexlab{c}}).

\bibitem[{\citenamefont{Wilson}(1974)}]{W7445}
\bibinfo{author}{\bibfnamefont{K.~G.} \bibnamefont{Wilson}},
  \bibinfo{journal}{Phys. Rev. D} \textbf{\bibinfo{volume}{10}},
  \bibinfo{pages}{2445} (\bibinfo{year}{1974}).

\end{thebibliography}

\end{document}